\documentclass[aps,prc,twocolumn]{revtex4-2}

\usepackage[english]{babel}
\usepackage{indentfirst} 
\usepackage[latin1,utf8]{inputenc}
\usepackage{lmodern}
\usepackage[T1]{fontenc}
\usepackage[pdftex]{graphicx}
\usepackage{setspace}
\usepackage{hyperref}

\usepackage{cellspace} 
\usepackage{amsmath,amsfonts,amssymb}

\begin{document}

\title{Strangeness production in the new version of the Liège Intra-Nuclear Cascade model}

 \author{J. Hirtz$^{1,2}$}
 \author{J.-C. David$^{1}$}
 \author{A. Boudard$^{1}$}
 \author{J. Cugnon$^{3}$}
 \author{S. Leray$^{1}$}
 \author{I. Leya$^{2}$}
 \author{J. L. Rodríguez-Sánchez$^{1,4,5}$}
 \author{G. Schnabel$^{1}$}
 
 \affiliation{$^1$IRFU, CEA, Université Paris-Saclay, F-91191, Gif-sur-Yvette, France}
 \affiliation{$^2$Space Research and Planetary Sciences, Physics Institute, University of Bern, Sidlerstrasse 5, 3012 Bern, Switzerland}
 \affiliation{$^3$AGO department, University of Liège, allée du 6 août 17, bât. B5, B-4000 Liège 1, Belgium}
 \affiliation{$^4$Universidad de Santiago de Compostela, E-15782 Santiago de Compostela, Spain}
 \affiliation{$^5$GSI (GSI Helmholtz Centre for Heavy Ion Research GmbH, D-64291 Darmstadt, Germany}

\begin{abstract}
The capabilities of the new version of the Liège Intra-Nuclear Cascade model (INCL++6) are presented in detail. This new version INCL is able to handle strange particles, such as kaons and the $\Lambda$ particle, and the associated reactions and also allows extending nucleon-nucleon collisions up to about $15-20$~GeV incident energy. Compared to the previous version, new observables can be studied, \textit{e.g.}, kaon, hyperon, and hypernuclei production cross sections (with the use of a suitable de-excitation code) as well as aspects of kaon-induced spallation reactions. The main purpose of this paper is to present the specific ingredients of the new INCL version and its new features, notably the new variance reduction scheme. We also compare for some illustrative strangeness production cases calculated using this version of INCL with experimental data.
\end{abstract}

\maketitle

\section{Introduction}
\label{intro}

Spallation reactions have applications in many different domains, like medicine, astrophysics, nuclear physics, etc. This large range of applications explains why it was, and still is, so important to fully understand and model this type of spallation reactions as stand-alone models or embedded in transport codes. Spallation models are commonly used in the energy range $100$~MeV-$3$~GeV, the higher energies being treated with string models. However, spallation models can in principle be used in a wider energy range. Notably, they can be used at energies up to $15-20$~GeV if the new open channels are correctly treated.

The Intra-Nuclear Cascade model of Liège (INCL), which was recognised by the IAEA benchmark in 2010 \cite{iaea} as one of the best spallation models when combined to the ABLA07 \cite{abla} de-excitation code, has been considerably improved during the last decade. One can mention the light ion induced reactions \cite{light.ion2}, the improvements for  few nucleons removal \citep{removal,jose-remove}, and the extension to high energy \citep{Pedoux,jc}. The last point was motivated by various reasons. A notable example is the need to make a better transition between string models like the FTF model \cite{fritiof}, for which the few GeV scale represents the low-energy part, and Intra-Nuclear Cascade (INC) models like INCL where the few GeV scale represents the high-energy part. This energy range is crucial notably for cosmic rays interactions. We can also mention an interest in exotic physics, like processes involving strangeness that appear in this new region with new particles and hypernuclei formation.

The high energy improvement of INCL involved three successive stages: multipion production in the final state of binary collisions \cite{Pedoux}, production of $\eta$ and $\omega$ particles \cite{jc}, and production of strange particles described for the first time in this paper. This last step is based on a previous work, which reunified the main ingredients needed for the implementation of strange particles \cite{bibi}. Here, we present the new version of INCL and the calculation results related to strange particle production. The observables, which will be discussed, depend only on the intra-nuclear cascade. The observables depending on both the intra-nuclear cascade and the de-excitation model will be discussed in a future publication.

The paper is organised as follows. In \autoref{present}, we present the main changes in INCL since the previous public version and in \autoref{biasing} we explain in detail the variance reduction methods introduced in INCL in order to facilitate the study of strangeness production physics. Next, in \autoref{results}, we compare the results obtained with the new version of INCL to experimental data of strange particle production for different initial conditions. Finally, \autoref{conc} is devoted to the conclusions and perspectives.

\section{The present INCL version}
\label{present}

The INCL++ version described in ref.\cite{light.ion2}, has essentially the same physics content, concerning nucleon-nucleus and pion-nucleus interactions, as the Fortran version INCL4.6 (\citep{incl4.6} and references therein). Only minor differences exist. However, one relevant difference between the INCL++ and the Fortran version is the capability to treat reactions induced by light ions ($A \leq 18$). Since the publication of ref.~\cite{light.ion2}, two main aspects were improved: the few nucleons removal channels, which are discussed in details in ref.~\citep{removal,jose-remove} and the high energy part \citep{Pedoux,jc,bibi}.

The extension of INCL to high energies allows simulations up to $\sim15$~GeV instead of the former upper limit of $2-3$~GeV of the previous version. This high energy extension was performed in three steps: multipions emission in binary collisions, which is the main contribution at high energies \cite{Pedoux}; implementation of $\eta$ and $\omega$ particles \cite{jc}; and finally, implementation of kaons and hyperons, described for the first time in this paper. In this section, we describe the different inputs and implementations linked to processes involving strageness. Cross sections are crucial ingredients. Although they are discussed below, the reader will find all details in ref.~\cite{bibi}.

The implementation of strangeness was divided into four main parts: first, the new particles to implement were selected. Then, their characteristics (mass, charge, isospin, decay modes, and average nuclear potential) were provided. Third, the interactions of these particles were described essentially with the use of experimental measurements and isospin symmetry. This includes the determination of production, scattering, and absorption cross sections as well as the phase phase generation of outgoing particles in binary collisions involving strange particles in the initial state and/or in the final state. The last part was the post-cascade treatment of the new particles.

\subsection{Particles}

The very first step before strangeness could be implemented was to determine which particles are to be added in INCL. This was based on two criteria. The first obvious criterion is the production rate in $NN$ and $\pi N$ collisions, which are the most common reaction types involved in intranuclear cascades. However, we also considered the quantity of available information in addition to a priori knowledge. For those reasons, it was decided to consider the kaons ($K^0$ and $K^+$), the antikaons ($\overline{K}^0$ and $K^-$), the $\Sigma$'s ($\Sigma^-$,$\Sigma^0$ and $\Sigma^+$), and the $\Lambda$. The difference between kaons and antikaons is relevant in this paper because the opposite hypercharge leads to significant differences, notably in the production, scattering, and absorption reactions. The heavier strange particles were not considered, due to lower production rates (they might be added in future extensions). The $\Phi$ meson ($s\overline{s}$) was not explicitly taken into account due to the lack of experimental data, even if it may play a role in the kaon-antikaon pair production. Some of its contributions are, however, hidden in the cross sections (as it is the case for other resonant particles).

\subsection{Average nuclear potentials and characteristics}

The mass, charge, isospin, decay modes, and half-life of each considered particle are taken from the Particle Data Group review \cite{pdg}. The average nuclear potential is much more difficult to determine. Only some of the implemented particles are relatively well known ($\Lambda$, $K^+$, and $K^-$). Studies on the $\Sigma$'s potential are rather sparse and it is not yet clear whether the $\Sigma$'s potential is attractive or repulsive. The $K^0$ and $\overline{K}^0$ are also problematic regarding their nature with the propagation eigenstates being different to the interaction eigenstates. 

Because few experimental measurements exist, the average potentials for strange particles in INCL are considered constant for sake of simplicity. In the case of the $\Lambda$ particle, a dependence on the nuclear asymmetry \cite{jose} is introduced to improve hypernuclei physics. Typically, the potential used for the $\Lambda$ is a $28$~MeV attractive potential for the symmetric nuclei that grows up to a $41$~MeV attractive potential for the highest asymmetries ($(A-2~Z)/A = 0.25$). The $K^+$ and $K^-$ have been relatively well studied and it is commonly accepted that the $K^+$ potential is slightly repulsive and the $K^-$ is strongly attractive \cite{pot-meson}. The values retained in this paper are a $25$~MeV repulsive potential for the $K^+$ and a $60$~MeV attractive potential for the $K^-$. It was decided to consider the same potential for $K^0$ and $\overline{K}^0$ as for $K^+$ and $K^-$ respectively with a slight correction ($10$~MeV) due to Coulomb repulsion. This choice is consistent with experimental measurements summarised in \cite{pot-meson}. The potential for $\Sigma$ particles is extremely difficult to measure because of nuclear effects. A repulsive potential of $16$~MeV for all $\Sigma$'s is used in INCL based on a recent study \cite{pot-sigma}.

In summary, the potentials used in this new version of INCL are:
\begin{equation}
    \left\{
      \begin{aligned}
         &25~MeV, \qquad &K&^+, \\
         &15~MeV,   &K&^0 , \\
         -&60~MeV,  &K&^- , \\
         -&50~MeV,  &\overline{K}&^{0} , \\
         &16~MeV,   &\Sigma&'s, \\
         &[-28,-41]~MeV,  &\Lambda&~(\text{Ref.\cite{jose}}). \\
      \end{aligned}
    \right.
\end{equation}

The difference between the potentials for kaons and antikaons potentials is not considered in other codes that we use for our comparisons. This produces significant differences in cross section prediction near threshold energies.

\subsection{Interactions}

As mentioned previously, the methods used to determine production, scattering, and absorption cross sections and the angular distributions of outgoing particles are described in detail in ref.~\cite{bibi}. In \autoref{reac1}, we summarise the considered reactions based on experimental data.

\begin{table}[ht]
\begin{center}
\begin{tabular}{ccl|ccl}
$NN$ & $ \rightarrow$ & $ N \Lambda K$        & $N \overline{K}$ & $ \rightarrow$ & $ N \overline{K}$          \\
     & $ \rightarrow$ & $ N \Sigma K $        &                  & $ \rightarrow$ & $ \Lambda \pi$             \\
     & $ \rightarrow$ & $ N \Lambda K \pi$    &                  & $ \rightarrow$ & $ \Sigma \pi$              \\
     & $ \rightarrow$ & $ N \Sigma K \pi$     &                  & $ \rightarrow$ & $ N \overline{K} \pi$      \\
     & $ \rightarrow$ & $ N \Lambda K \pi\pi$ &                  & $ \rightarrow$ & $ \Lambda \pi \pi$         \\
     & $ \rightarrow$ & $ N \Sigma K \pi\pi$  &                  & $ \rightarrow$ & $ \Sigma \pi \pi$          \\
     & $ \rightarrow$ & $ NN K \overline{K}$  &                  & $ \rightarrow$ & $ N \overline{K} \pi \pi$  \\
\\
$\pi N$ & $ \rightarrow$ & $\Lambda K $       & $N K$        & $ \rightarrow$ & $ N K $        \\
        & $ \rightarrow$ & $\Sigma K $        &              & $ \rightarrow$ & $ N K \pi$     \\  
        & $ \rightarrow$ & $\Lambda K \pi$    &              & $ \rightarrow$ & $ N K \pi \pi$ \\
        & $ \rightarrow$ & $\Sigma K \pi$     & $N \Lambda $ & $ \rightarrow$ & $ N \Lambda$   \\
        & $ \rightarrow$ & $\Lambda K \pi\pi$ &              & $ \rightarrow$ & $ N \Sigma$    \\
        & $ \rightarrow$ & $\Sigma K \pi \pi$ & $N \Sigma $  & $ \rightarrow$ & $ N \Lambda$   \\
        & $ \rightarrow$ & $N K \overline{K}$ &              & $ \rightarrow$ & $ N \Sigma$
\end{tabular}
\caption{\label{reac1} List of considered reactions involving strangeness based on experimental data.}
\end{center}
\end{table}

In addition to the reactions listed in \autoref{reac1}, we include two other types of reactions, which are listed in \autoref{reac2}. Firstly are the $\Delta N$ reactions, which are expected to contribute significantly to strangeness production according to the study of Tsushima et \textit{al.}\cite{tsushima}. The second type is the strangeness production in reactions where many particles are produced in the final state but no measurements are available. This second type is needed to get the correct inclusive strangeness production cross section. Since the cross sections given in \autoref{reac2} are coming from models, larger uncertainties are expected.

\begin{table}
\begin{center}
\begin{tabular}{ccl|ccl}
$\Delta N$ & $ \rightarrow$ & $N \Lambda K$         & $NN$   & $ \rightarrow$ & $K + X$  \\
           & $ \rightarrow$ & $N \Sigma K $         &        &                &          \\  
           & $ \rightarrow$ & $ \Delta \Lambda K$   &$\pi N$ & $ \rightarrow$ & $K + X$  \\
           & $ \rightarrow$ & $ \Delta \Sigma K$    &        &                &          \\
           & $ \rightarrow$ & $ NN K \overline{K}$  &        &                &          \\
\end{tabular}
\caption{\label{reac2} List of the reactions involving strangeness and requiring information that has been taken exclusively from models. The $X$ stands for all possible reactions excluding the reaction summarised in \autoref{reac1}}
\end{center}
\end{table}

\subsection{Post-cascade treatment}

The implementation of strangeness can lead to a new situation at the end of the intra-nuclear cascade; a hyperremnant can be produced in which at least one strange particle is still inside the target nucleus at the end of the intra-nuclear cascade. Therefore, we had to decide how to treat the remaining strange particle after the end of the cascade.

Owing to the repulsive average nuclear potential for the kaons, we decided to eject the trapped kaons at the end of the cascade and to correct their kinematics accordingly to their potential.

All $\Sigma$'s and antikaons have high absorption cross sections ($N \Sigma \rightarrow N \Lambda$ and $N \overline{K} \rightarrow \Lambda \pi$) at low energy. Therefore, we decided to absorb all of them when they are trapped inside the nucleus and to convert the excess of energy and the mass energy of the possible pions into nucleus excitation energy.

After kaon emission and $\Sigma$ and antikaon absorption, the hyperremnant contains only protons, neutrons, and $\Lambda$'s. The hyperremnant is then de-excited using a new version of the ABLA07 code \cite{abla}, which will be presented in a future paper.

\section{Variance reduction methods}
\label{biasing}

Strangeness production is a rare process in spallation reactions, especially for energies below $5-6$~GeV. However, the user of INC models might have an interest in strangeness physics. Therefore, in order to have enough statistics, a calculation will require a large number of events, an event being defined as the simulation of one cascade in INCL. This would take a lot of computational time and result in a large amount of useless information; taking a lot of space on a hard drive and slowing down the analysis.

The solution proposed here is to use variance reduction methods, which are based on the importance sampling method. In INCL++6, we implemented a variance reduction scheme (VRS), which artificially increases the statistics for the observables linked to the strangeness production, keeping the calculation time and the output file size unchanged.

Note that INCL has an option enabling to write into the final ROOT file only if a certain condition is fulfilled. This makes the output size problem marginal for us. However, the information that does not fulfil the condition is lost with this option. With the new VRS all the information is kept, which we consider a major benefit.

This section describes the variance reduction scheme and its operation. The reader only interested in the calculation results of INCL dealing with strangeness production can skip this section. The variance reduction method developed here, and used in INCL, aims at getting results with lower uncertainties, and in some cases is the only way to get results within a reasonable computation time, but it does not change the conclusions. Readers eager to understand the method will find in the following some generalities (\autoref{Gen}) presenting the topic, the constraints and the difficulties within INCL, the scheme developed, and the technical aspects (\autoref{reversal_sec}), the case of correlations (\autoref{eir}), the reliability of the method (\autoref{vali}), and some examples of uses of the VRS (\autoref{examples}). Readers interested by the main lines and aspects, and not by the method itself, can go directly to \autoref{vali} and \autoref{examples}.

\subsection{Generalities}
\label{Gen}

Basically, a fully operational important sampling method consists of two successive steps. In a first step, the rules of the simulation, \textit{i.e.}, the physics ingredients, are modified (biased) in order to improve the sampling. In our approach, we increase the strangeness production. In a second step, the simulation result is accordingly corrected. The goal of the variance reduction methods is to obtain the true observable of interest (\textit{e.g.}, a cross section) with a reduced variance and therefore with reduced uncertainties within the same computational time.

For this study, all the processes associated with strangeness production are of interest. Therefore, the VRS of INCL was developed to increase the realisation probability for all of them, independently of the observable. However, even though the current VRS globally modifies the probabilities for processes involving strangeness, other schemes are in principle possible. The scheme could also be adapted to be more restrictive (\textit{e.g.}, forward kaon production) or to work for another particle type (\textit{e.g.}, $\eta$ meson production) or physics (\textit{e.g.}, peripheral collisions).

The reason for the low global strangeness production rates in INCL is the low elementary strangeness production cross sections. For example, the strangeness production represents 0.014\%(0.15\%) of the total cross section in proton-proton collision at kinetic energies of $2(3)$~GeV. Therefore, in a first step, the VRS modifies the hadron-hadron cross sections in order to increase the strangeness production. In a second step, it corrects a posteriori the bias introduced in order to obtain the unbiased observable estimators with reduced uncertainties.

In the general case, the second step (\textit{i.e.}, the bias reversal) consists in determining the ratio of the probability to make an observation in the non-biased version to the probability to make the same observation in the biased version. This ratio is called the importance or the weight and can be calculated for a complete event (a cascade), a specific particle or for any observation during the event. Next, the importance is used to weight the contribution of observations. This number gives the information about how much this observable is biased. The importance for an observable $X$ can be written as:
\begin{equation}
\label{br}
W_X = \frac{P(X|\text{no-bias})}{P(X|\text{bias})}.
\end{equation}

This expression of the importance leads to in a first requirement that events contributing to a variable of interest should have a non-zero probability of realisation in the biased version (the version using the variance reduction scheme) otherwise it will result in an arithmetic exception. In other words, every strange event in INCL that can be produced in the standard version must be attainable in the version using the variance reduction scheme. Only the probability of realisation can be changed. Because of this constraint and because it is not trivial to know whether strangeness production could occur later during an event, no channel cross section can be reduced to zero at the binary collision level in INCL.


The treatment of a particle during an INCL event can be subdivided into a three step cycle. First, the particles are propagated inside the nucleus. This step ends when two particles collide. A collision occurs when the distance between two particles is below a maximal interaction distance based on their total interaction cross sections. Second, the type of the reaction of the binary collision is randomly chosen based on the respective reaction cross sections. In the last step, the phase space and the charge repartition is randomly generated either based on differential cross sections (if available) or on phase space generators. Then, the cycle is repeated until the end of the intra-nuclear cascade.

The propagation along straight trajectories between collision events is deterministic. According to this, an artificially decreased of a total interaction cross section can lead to a situation of particles flying past each other where they would have collided using the original cross section. From such an event onwards, the subsequent cascade is outside the universe of possibilities based on the unbiased total interaction cross section. Therefore, the importance of this event would be null according to \autoref{br}. The same argument can be made for an increased total interaction cross section. Hence, the total interaction cross sections must be conserved.

Thanks to the random treatment of the reaction choice and the phase space generation, both steps can be biased; changing the probability of realisation. In our case, only the step selecting the type of reaction needs to be biased to increase the global strangeness production. However, if the user is interested in the production of a particle in a specific phase space (\textit{e.g.}, backward production), phase space generation can be biased under minor modification in the code.

The two constraints, the prohibition to cancel a channel and the total cross section conservation, will be crucial for the INCL variance reduction scheme.

With the variance reduction scheme, the new weighted estimators and the associated variances of an observable $X$ \cite{owen} are given by:
\begin{equation}
\label{hope}
E(X) = \frac{\sum_{i=1}^M w_i x_i}{N},
\end{equation}
\begin{equation}
\label{var}
V(X) = \frac{\sum_{i=1}^M (w_i x_i - E(X))^2}{M},
\end{equation}
with $w_i$ the importance of the $i^{th}$ observation $x_i$ of the observable $X$ and $N$ the number of events. The natural summation is such case is when summing over the events ($M=N$), with $w_i$ the importance of the $i^{th}$ event and $x_i$ the number of observations corresponding to the observable $X$ (\textit{e.g.}, kaons) in the corresponding event.

An alternative summation is actually possible and preferred in our case. We can sum over the final particles ($M \neq N$) with $w_i$ the importance of the $i^{th}$ particle and $x_i=1$ or $0$ if the particle corresponds to the observable or not, respectively. With the natural summation, we estimate the mean value of the number of strange particles produced per reaction. In the alternative summation, we estimate the number of strange particles produced then we normalise by the number of reactions. It clearly appears the observables estimated correspond in the two approaches.

For INCL, it is simpler to sum over the particles considering the output generated. Moreover, for sake of simplicity, the relative uncertainties displayed in further figures do not derivate from \autoref{var} but are calculated using the formula:
\begin{equation}
\label{relunc}
relat.~uncer. = \frac{\sqrt{\sum (w_i x_i)^2}}{\sum w_i x_i},
\end{equation}
which is equivalent to the standard relative uncertainty equals to the inverse square root of the number of observations in the Monte Carlos simulations and which is an approximation of \autoref{var}.

It is worth emphasising that, the observable $X$ can be anything. For example, it can be the presence or absence of a particle of a certain type, the number of particles of a certain type produced during the cascade, the presence of many particle correlated, an entire event, or a part of an event.

The minimum variance for a given number of events is achieved when all the strange particle importances are equal according to \hyperref[hope]{Equations~\ref*{hope}} and \ref{var}. However, these importances are not always equal in INCL for reasons explained further below. Therefore, the objective is to keep them as close as possible to minimise the variance and therefore uncertainties.

\subsection{Variance reduction scheme in INCL}
\label{reversal_sec}

\begin{figure}
\centering
\includegraphics[width=\columnwidth]{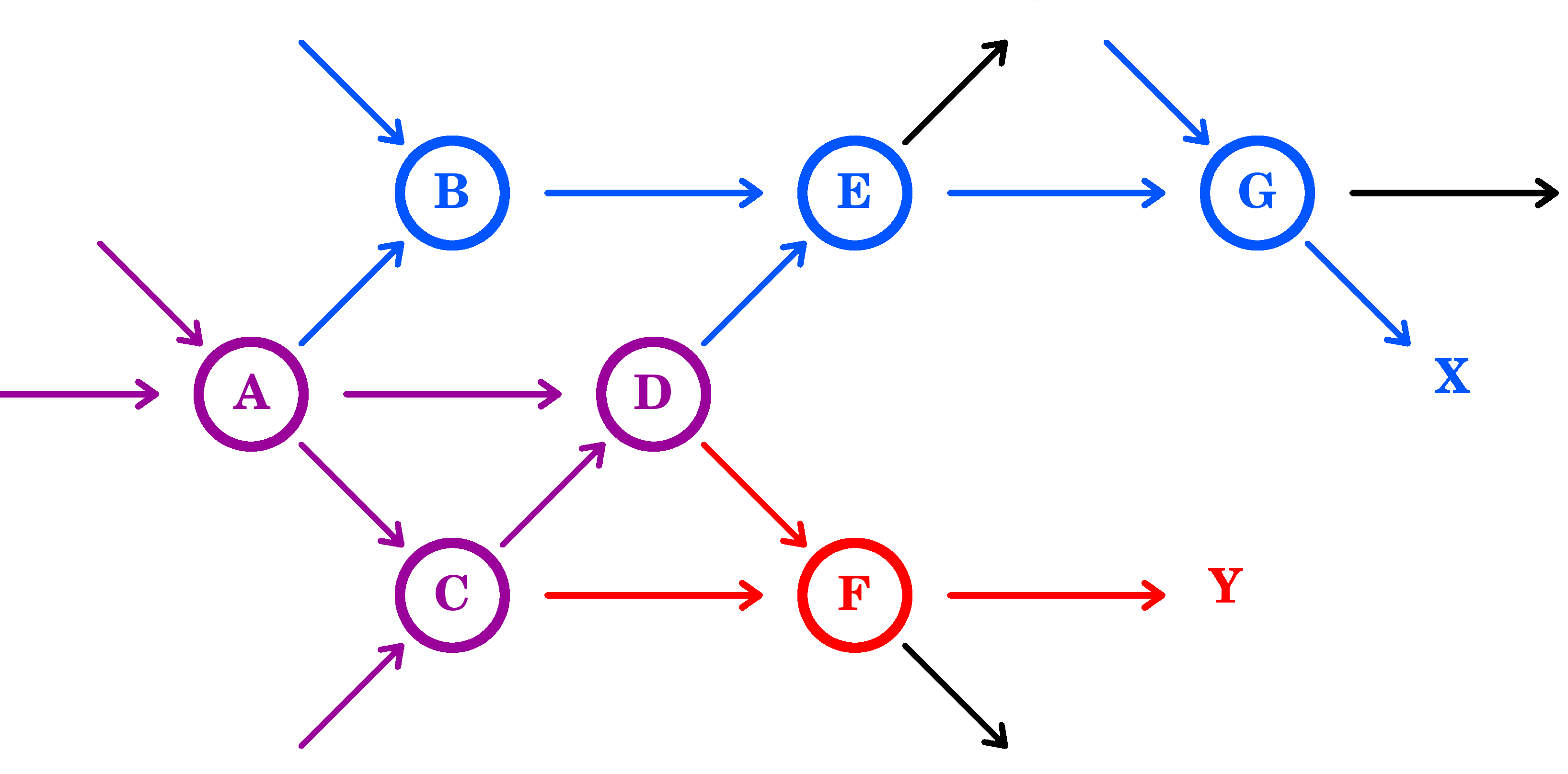}
\caption{\label{graph} Simplified example of an intra-nuclear cascade represented as a time ordered graph. Circles and arrows represent binary collisions and the propagation of the particles, respectively. The blue (red) and purple part of the graph is the history of the $X$ ($Y$) particle.}
\end{figure}

For a better understanding of the VRS used with INCL, a cascade can be seen as a time ordered graph where the edges (arrows) and vertices correspond to the particles propagating and to the binary collisions, respectively. Thus, a cascade is fully defined by the set of its vertices. The information about what happened before (\textit{e.g.}, the projectile type, the impact parameter, ...) is hidden in the definition of the vertices initial state. A schematic example is displayed in \autoref{graph}.

As explained in the previous subsection, the VRS must increase the strangeness production in a way that minimises the spread of the strange particle importances to achieve minimum variance and therefore to obtain the smallest possible uncertainties for a given number of events.

In our variance reduction scheme, if the particle $Y$ shown in  \autoref{graph} is a strange particle, the event containing the set of vertices $A$, $C$, $D$, and $F$ should have an increased probability of realisation. If the particle $X$ is a strange particle, the events containing the set of vertices $A$, $B$, $C$, $D$, $E$, and $G$ should be promoted. If both particles are strange particles, both paths should be promoted in the same way. Two solutions can be considered. The first one is to promote the vertices $A$, $C$, and $D$, which are common to both paths, and to continue as a standard cascade. In such an approach, it actually means only the first vertex can be biased. The problem is that we do not know whether if a strange particles will be produced in any of the binary collision represented by a later vertex. As a consequence, only strangeness production in the first collision can be promoted. This would lead to differences between the importances for the events that produced strangeness in the first collision, and the events that produced strangeness in secondary collisions. Thus, is would be a new source of variance, what should be avoided. The second solution is to bias each vertex (each elementary binary collision) along the entire cascade in such a way the same importance is obtained for every strange particle. The second solution was chosen for INCL and the importance associated to each final particle is provided in the final ROOT file.

In our solution, the promotion of strangeness production can change from one binary collision to another. However, if different channels are producing strangeness in a given binary collision, they should be promoted in the same way in order to get the same importance as the other strange particles whatever the channel chosen. Once again, this should be done in order to minimise the variance.

With the INCL variance reduction scheme, we introduce a new input parameter to INCL. This input is a scalar, which is used to defined the wanted importance of strange particles. This allows to have the same importance for strange particles in the different events, which are independent. This input scalar will be called the \textit{bias factor} in the following. The VRS implemented in INCL tries to force the final importances of strange particles to be the reciprocal value of the bias factor. Consequently, the bias factor is a multiplication factor of the probability to create a strange particle. However, the effective increase is lower in some cases. This is discussed in \autoref{examples} with the way to optimise the bias factor.

Before explaining how the VRS forces the final importances of strange particles to be a specific value, we describe how the importances of particles are calculated. Doing so, we start with the general formula:
\begin{equation}
\label{base}
P(N \cap M) = P(N|M) \times P(M),
\end{equation}
with $P(N \cap M)$ the probability of realisation of $N$ and $M$, $P(N|M)$ the probability of realisation of $N$ knowing that $M$ is realised, and $P(M)$ the probability of realisation of $M$.

In \autoref{graph} the vertex $B$ can happen only if the vertex $A$ has been realised, therefore:
\begin{equation}
\label{logic}
P(A|B) = 1.
\end{equation}

Combining \hyperref[base]{Equations~\ref*{base}} and \ref{logic} we get:

\begin{equation}
P(A \cap B) = P(A|B) \times P(B) = P(B).
\end{equation}

For the same reason:
\begin{gather}
P(A|F) = P(C|F) = P(D|F) = 1,\\
\Rightarrow P(A \cap C \cap D \cap F) = P(F).
\end{gather}

Therefore, in the version using the VRS the probability of producing the final particle $Y$, which is produced in vertex $F$, must be corrected by the importance $W_Y$, which is given by:
\begin{align}
\label{eq_w_y}
W_Y & = W_F  = \frac{P(F|\text{no-bias})}{P(F|\text{bias})}, \nonumber \\
& = \frac{P(A \cap C \cap D \cap F|\text{no-bias})}{P(A \cap C \cap D \cap F|\text{bias})}, \nonumber \\
& = \frac{P(A|\text{no-bias})}{P(A|\text{bias})} \times \frac{P(C|A,\text{no-bias})}{P(C|A,\text{bias})} \nonumber  \\
& \ \times \frac{P(D|A,C,\text{no-bias})}{P(D|A,C,\text{bias})} \times \frac{P(F|A,C,D,\text{no-bias})}{P(F|A,C,D,\text{bias})}, \nonumber \\
& = \frac{\sigma(A)/\sigma_{tot}(A)}{\sigma'(A)/\sigma_{tot}(A)} \times \frac{\sigma(C)/\sigma_{tot}(C)}{\sigma'(C)/\sigma_{tot}(C)} \\
& \ \times \frac{\sigma(D)/\sigma_{tot}(D)}{\sigma'(D)/\sigma_{tot}(D)} \times \frac{\sigma(F)/\sigma_{tot}(F)}{\sigma'(F)/\sigma_{tot}(F)}, \nonumber \\
& = \frac{\sigma(A)}{\sigma'(A)} \times \frac{\sigma(C)}{\sigma'(C)} \times \frac{\sigma(D)}{\sigma'(D)} \times \frac{\sigma(F)}{\sigma'(F)}, \nonumber \\
& = CSR(A) \times CSR(C) \times CSR(D) \times CSR(F), \nonumber
\end{align}
with $CSR(I)$ the cross section ratio of vertex $I$, $\sigma(I)$ and $\sigma'(I)$ the standard and biased cross section of the reaction that took place in vertex $I$, and $\sigma_{tot}(I)$ the total interaction cross for the vertex $I$.
 $W_Y = W_F$ because the propagation is not biased.

The cross section ratios are easily determined during an event since both terms in the ratio $\sigma/\sigma'$ are known for all the possible reactions when a binary collision happens. Some examples are discussed in \hyperref[importance_ratio]{appendix~\ref*{importance_ratio}}. 

In the VRS of INCL, the cross section ratio of the vertices are stored. Therefore, whenever a binary collision happens, the cross section ratio of the previous vertices are known. This allows to calculate what should be the cross section ratio of the vertex representing the aforementioned binary collision (\textit{e.g.}, the vertex $F$ for the case in \autoref{eq_w_y}) to match the importance of the outgoing particles (the particle $Y$ in our example) to the desired importance. This calculated cross section ratio defines how the strangeness production should be promoted (see examples in \hyperref[importance_ratio]{appendix~\ref*{importance_ratio}}).

In the special case of a binary collision between a strange particle and another particle, a strange particle will be present in the final state regardless of the channel chosen in INCL. Therefore, the strangeness cannot be promoted and the cross section ratio will be equal to $1$. Such a binary collision can result in a strange particle that does not have the aimed importance of the strange particles. No solution was found to solve this problem. However, the dispersion of particle importances due to this phenomenon in most of cases is marginal thanks to the precautions discussed below and does not introduce a new significant source of variance.

The main source of importance dispersion for strange particles in INCL, which is itself source of variance, is due to the two constraints discussed before. First, the non-strange cross sections cannot be null but no lower bound is fixed. Therefore, when a reaction cross section is drastically reduced but the corresponding reaction is chosen anyway, the cross section ratio of the binary collision is extremely large. Second, the total cross section must remain unchanged. This fixes a lower bound for the cross section ratio of a vertex that produces strangeness (=$\sigma^{strange}$/$\sigma^{tot}$). Therefore, if the cross section ratio required for a vertex is below this limit, the effective strangeness promotion will not correspond to the one required.

The first constraint can lead to extremely large cross section ratios, which cannot be counterbalanced by the following ones because of the second constraint. If a strange particle is produced on such a path, it would result in a strange particle with a large importance compared to others and, therefore, it will strongly contribute to the observables. Such high contributions produced by single particles will make the convergence slower, lead to variance \textit{jumps} and, therefore, to large uncertainties for the observables (see \hyperref[examples]{subsection~\ref*{examples}}). If pronounced \textit{jumps} are seen, all the observables estimated with the corresponding INCL results should not be trusted. In such a case, the associated uncertainty estimated accordingly to \autoref{var} or \autoref{relunc} might be highly underestimated because of the bad sampling. This situation means the bias factor was chosen way too large and paths important for strangeness production were suppressed too much.

To avoid these problems, a compromise should be found between the increase of the statistics, obtained using high bias factors, and the maintain of equal importances for strange particles, which is most easily achieved using small bias factors.

Therefore, a safeguard was implemented in the variance reduction scheme. This safeguard aims at optimising the convergence efficiency for observables of interest by modifying the bias factor used. The safeguard does it by preventing the decrease of channel cross sections below the half of the initial cross sections. Therefore, a vertex cross section ratio cannot be higher than $2$. This strongly limits the product of vertex cross section ratios for a given history and it will be easier to counterbalance it. At the end of an event, this procedure strongly reduces the variance \textit{jumps} even if the bias factor is chosen too high. A less restrictive safeguard has been tested but the actual one presents better compromises. However, this safeguard is not perfect. If a particle has a history with numerous vertices with cross section ratios between $1$ and $2$, it can have a large importance anyway and will result in variance \textit{jumps} and a slow convergence.

\subsection{Event importance reversal}
\label{eir}

As previously mentioned, \autoref{hope} is used to estimate the observables. Two types of summation can be used: a sum over the events using the importance of the events where $x_i$ is an integer (not used in our case) or a sum over the particles using the importance of the particles where $x_i=0$ or $1$. However, the dynamic adjustment of the bias factor depending on the history of events and their importances introduces some sort of dependency between particles of a same event. If working at the level of particle importances as we do, \autoref{hope} is probably not correct any more when correlations must be taken into account (\textit{e.g.}, when looking at cross section of hyperon emission in coincidence with a kaon). 

There is an alternative possible type of summation for \autoref{hope}. The summation can be done over the particles ($x_i=0$ or $1$) but using the event importances instead of the individual particle importances.

The event importance is equal to the product of every vertex cross section ratio of the entire event. It is, a priori, different from the importance of an observation $X$ (let us say a particle) because it includes the contribution of extra vertices, which are not on the path of the particle $X$, and are therefore not relevant. However, the expected value of the cross section ratio of a vertex $A$ is:
\begin{align}
\label{eq}
E(CSR(A)) & = \sum_{\text{reac}} \frac{\sigma( \text{reac})}{\sigma'(\text{reac})} \times P(\text{reac}|\text{init. state},\text{bias}) \nonumber, \\
& = \sum_{\text{reac}} \frac{\sigma(\text{reac})}{\sigma_{tot}} = 1,
\end{align}
with the same notation as in \autoref{eq_w_y}. Noteworthy, this equation is true only if no reaction of the non biased version is forbidden in the biased version, what is the case in INCL. The same argument can be used for any sub-structure in a cascade, which have no constraints on its final state. In particular, the extra vertices, which are not on the path of the particle $X$, have not constraint on their final state since they play no role in the production of the particle $X$. Therefore, the contribution of extra vertices is statistically null and the expected value of the event importance knowing the particle $X$ has been produced is equal to the importance of this particle $X$.

Then, it is trivial to prove to the expected value of the estimator is the same using $w_i$ ($= E(w_\text{event})$) or $w_\text{event}$ taking into account $x_i=0$ or $1$ in \autoref{hope} when using the summation over the particles.

\begin{figure*}[!t]
\begin{minipage}{2\columnwidth}
\begin{center}
\includegraphics[width=0.47\columnwidth ,keepaspectratio]{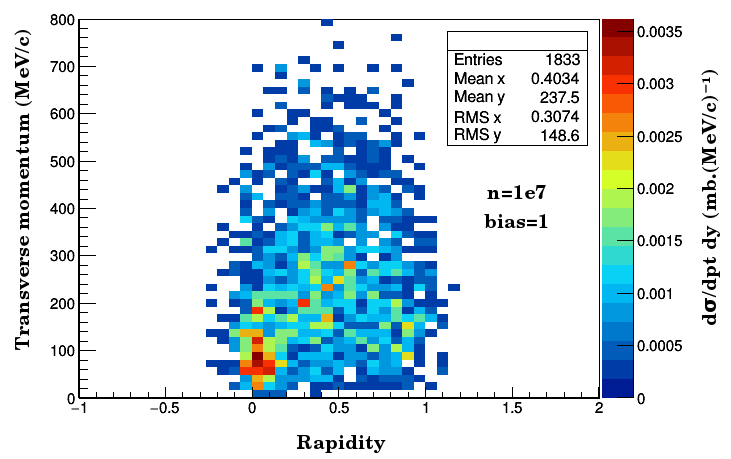} \hspace{3mm}
\includegraphics[width=0.47\columnwidth ,keepaspectratio]{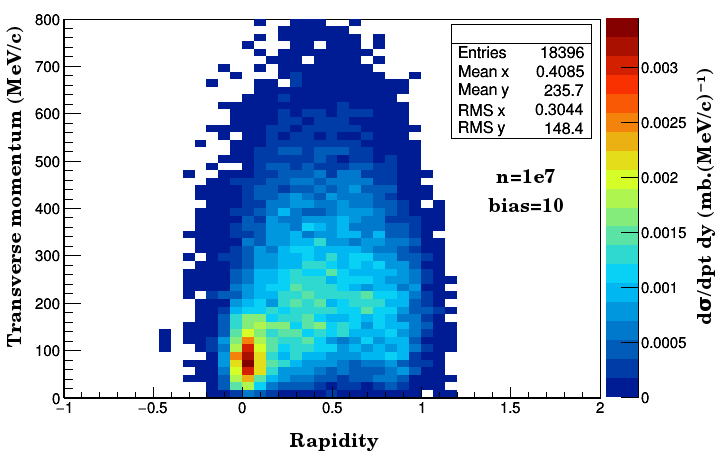}
\caption{\label{versus} $\Lambda$ transverse momentum versus rapidity distribution in $p(1.7~GeV)+Ca$ collision. Both plots are obtained using the same number of events ($10^7$). Left: no bias. Right: bias factor $= 10$.}
\end{center}
\end{minipage}
\end{figure*}

The event importance is provided in the final ROOT file. The convergence being slower or equal, it is recommended to use the particle importance for the reversal of independent observables. However, it was decided to conserve only a scalar importance for each particle and a event importance in order to limit the quantity of information recorded in the output file. Doing so, the correlations are lost. Therefore, if the observable is a coincidence production, the event importances must be used regardless of the type of summation. Additionally, if INCL is connected to another program (\textit{e.g.}, a transport code) that is not able to keep track of the importance of the individual particles, this program can use the event importance to weight its own results.

\subsection{Validation}
\label{vali}

As a first test of whether the VRS works reliable and as expected, the results obtained using the variance reduction method are compared with results obtained using the non-biased version. Not only the observables obtained with the VRS must be the same as the observables obtained without it (no residual bias), but the uncertainties must also be reduced for the same calculation time (variance reduction). Various representative comparisons are discussed in the following.

A comparison of simulations using or not the VRS indicates that the calculation time is independent of the bias factor used. Therefore, the number of events can be used as a measure of the computational time and/or to compare the efficiency of calculations using different bias factors.

\autoref{versus} shows a comparison of calculations with and without bias. The two calculations have identical inputs except for the bias factor ($1$ and $10$). It can be clearly seen that the calculation using the bias factor $=10$ (right panel) produced a much more precise distribution using the same binning thanks to the statistic increased by a factor around $10$. Additionally, no significant difference shows up for the amplitude, the mean values, and the standard deviations displayed for both bias factors in the frames. Consequently, the objectives of the VRS are perfectly fulfilled.

\begin{figure}[!b]
\centering
\includegraphics[width=\columnwidth]{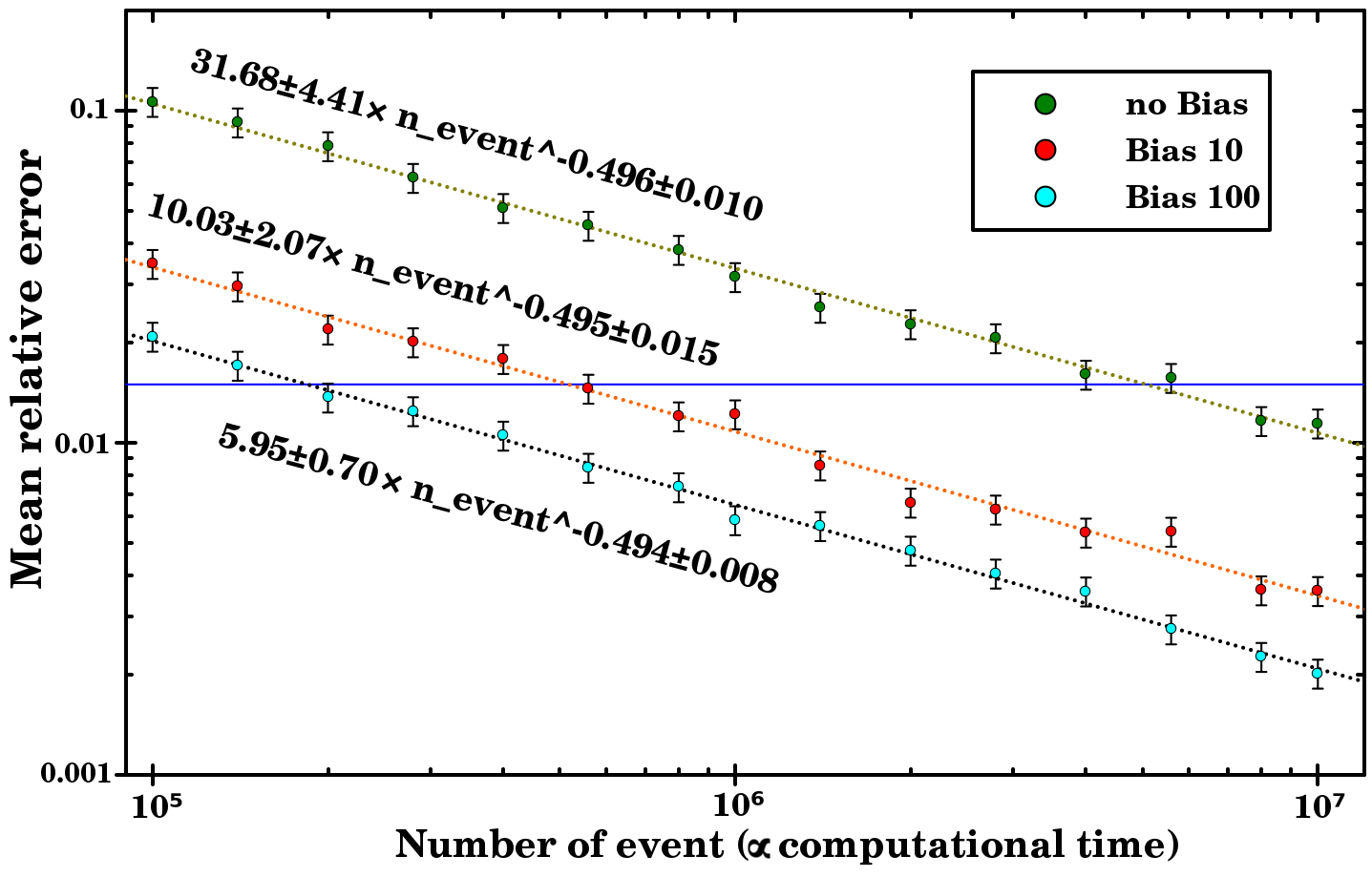}
\caption{\label{merit} Mean relative error of the $K^+$ mean momentum estimator as a function of the number of events. The number of events corresponds to a simulation time. The true value taken for the kaons mean momentum is estimated using a $10^9$ unbiased event calculation. The considered reaction is $p(1.7~GeV)+^{12}C$ with $bias~factor=10$~(red), $100$ (cyan) and not using a variance reduction method (green). Dotted lines are fits of the form $a~n^b$. The horizontal blue line is here to guide eyes (see text).}
\end{figure}

\begin{figure*}[t]
\begin{minipage}{2\columnwidth}
\begin{minipage}{0.47\columnwidth}
\includegraphics[width=\columnwidth]{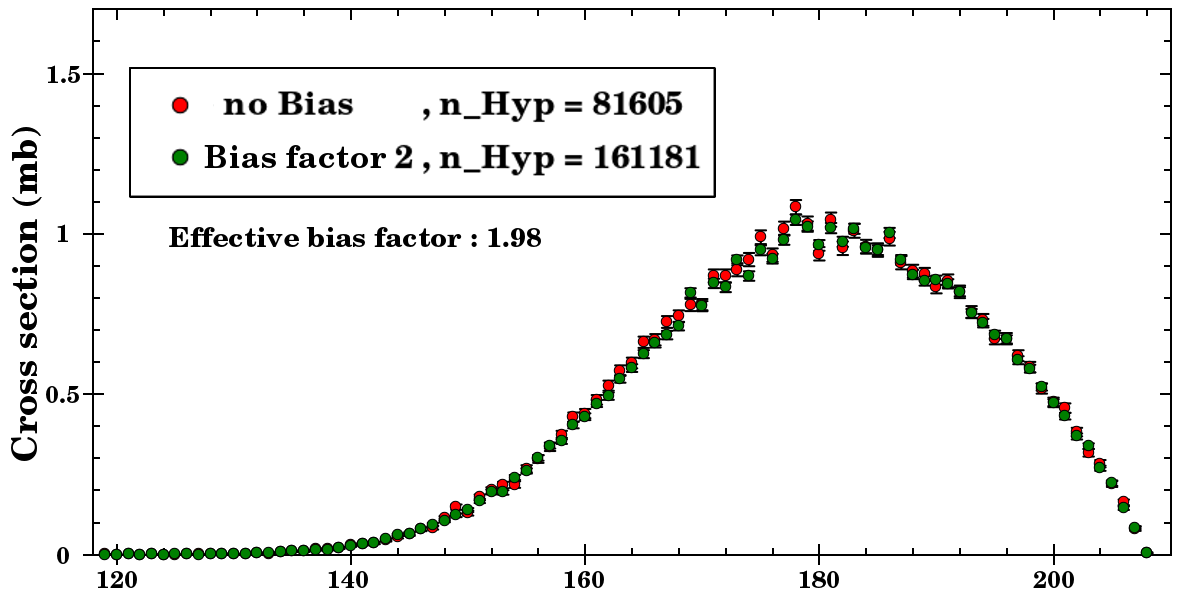}
\includegraphics[width=\columnwidth]{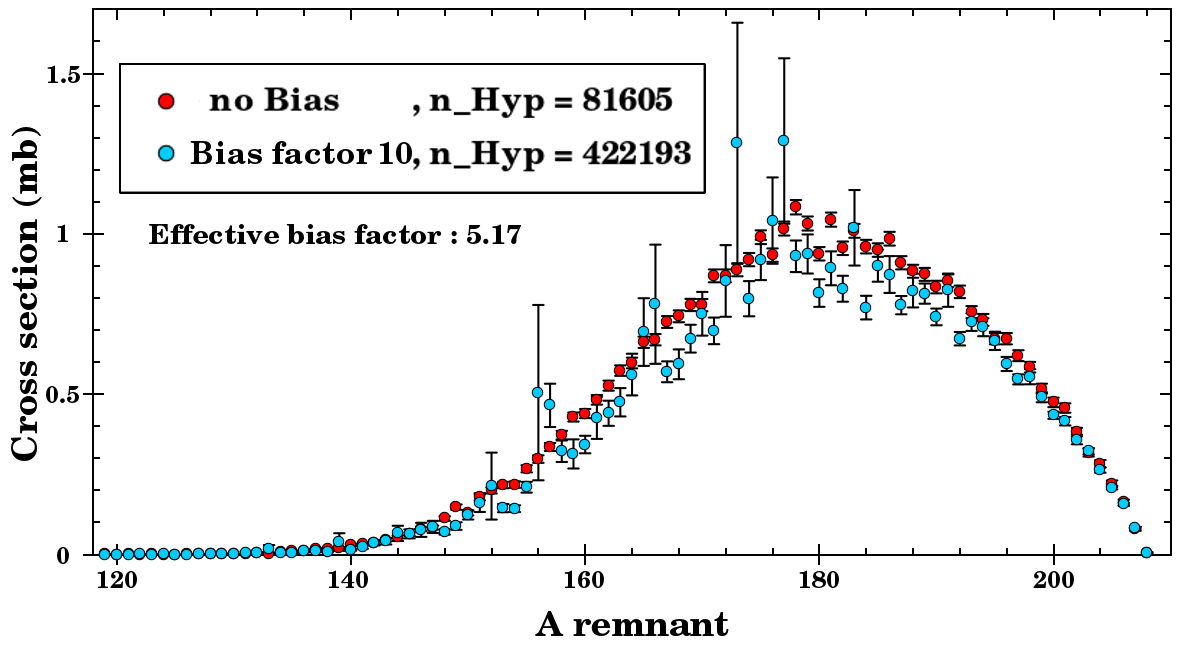}
\end{minipage}
\begin{minipage}{0.01\columnwidth}
~
\end{minipage}
\begin{minipage}{0.5\columnwidth}
\includegraphics[width=\columnwidth]{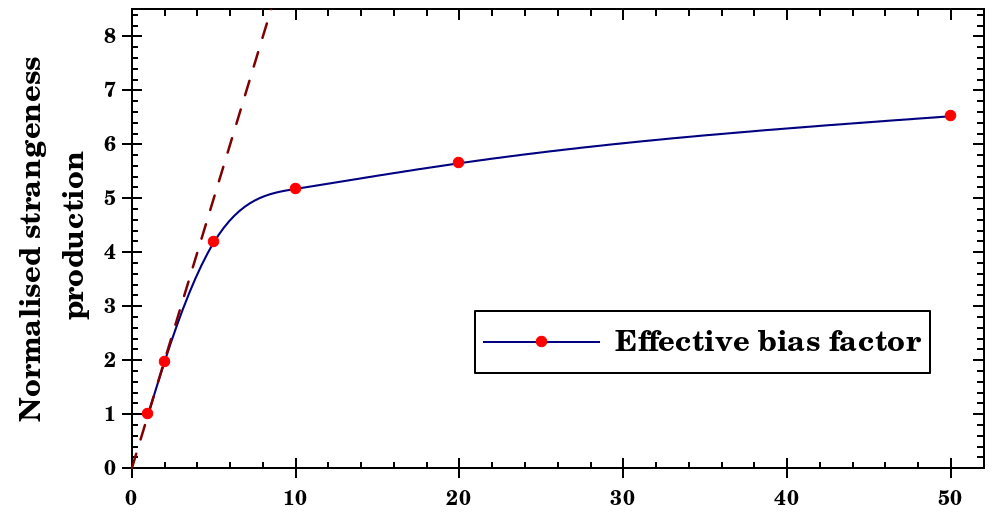}
\vspace{-7mm}
\begin{flushright}
\includegraphics[width=0.95\columnwidth]{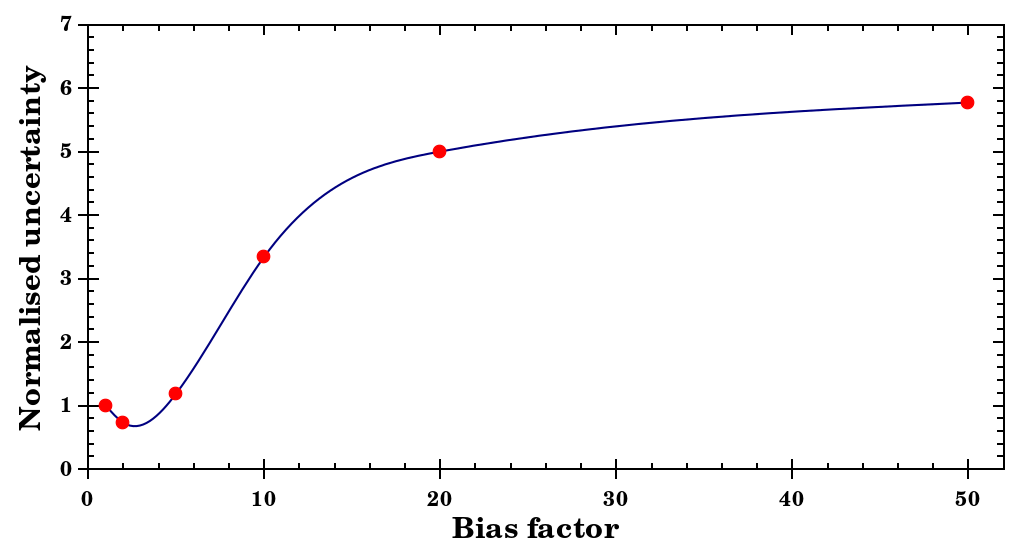}
\end{flushright}
\end{minipage}
\caption{\label{problem} Left: Hyperremnant mass distribution in $p(10~GeV)+^{208}Pb$ collisions with $10^7$ shots. Calculations with bias factors $=2$ (top) and $10$ (bottom) are compared to the calculation without VRS used. Right: Evolution of uncertainties and statistics as a function of the bias factor (see text). The uncertainties and statistics are normalised to $1$ for bias factor=$1$ (no VRS used). The dashed brown line in the top right panel represents the prefect case where the effective bias factor equals the nominal bias factor.}
\end{minipage}
\end{figure*}

A test of convergence was carried out for the reaction $p(1.7~GeV)+^{12}C$ using different bias factors (no bias, $10$, and $100$) using $4500$ simulations with various numbers of events per calculation. The observable chosen to test the convergence efficiency is the $K^+$ mean momentum. What is considered as the true value for the $K^+$ mean momentum was estimated with an additional calculation using no bias and $10^9$ events, which resulted in an uncertainty of $0.1\%$ for the true value. We then compared the estimator of the $K^+$ mean momentum for the $4500$ simulations to the true value. The absolute value of the difference between an estimator and the true value is the error. \autoref{merit} shows the mean relative error for the different bias factors as a function of the number of events, which corresponds to a computational time, which is independent on the bias factor. Fits of the form $a~n^b$ are plotted and the parameters and their uncertainties are shown.

Since the true value has been estimated with an uncertainty of $0.1\%$, a mean relative error of below $0.001$ cannot be interpreted in \autoref{merit}.

A horizontal line was added in \autoref{merit}. It represents a relative error of $1.5\%$. According to this, the precision of $1.5\%$ is obtained with about $5\times10^6$ events without the VRS and with only $5\times10^5 (1.8\times10^5)$ events with a bias factor $10(100)$. Therefore, the $1.5\%$ precision is reached with about $10$ times less events using a bias factor $10$, while using a bias factor of $100$ the gain in time is not $100$, but only $30$. This reduction of the computational time needed to reach the same precision is called the effective bias factor. The effective bias $30$ for the bias factor $100$ indicates that the optimal bias is probably around $30$ and the safeguard (see \autoref{reversal_sec}) forced the effective bias factor to be closer to this value.

In \autoref{merit}, a decreasing mean relative error that is inversely proportional to the square root of the number of events is observed for every bias factor used. This includes the bias factor that is clearly above the optimal bias factor. Additionally, the dispersion of the results for the $4500$ calculations follows a normal distribution around the true $K^+$ mean momentum (estimated using the calculation with $10^9$ events) with respect to their statistical uncertainties. This confirms that every calculation converges to the same limit regardless of the bias factor used. Therefore, the VRS (bias + reversal) does not introduce any bias in the final observables. They all converge to the true value.

\subsection{Examples}
\label{examples}

Let us remind the important nomenclature introduced in this section:
\begin{description}
\item[bias factor] The augmentation of the statistics of strange particles required at the beginning of the calculation.
\item[effective bias factor] The real augmentation of the statistics of strange particles observed.
\item[optimal bias factor] The bias factor minimising the uncertainties linked to strange particles for a given number of events.
\end{description}

The use of the VRS of INCL++6 requires two steps in addition to the standard use of INCL. Firstly, the user must provide the bias factor and second, the user must weight the final observables with the corresponding importances stored in the output file.

The a priori bias factor optimisation is not trivial because the optimal bias factor strongly depends on the initial parameters, like the target size and the projectile kinetic energy, but it depends also on the final observables. However, the bias factor does not need to be perfectly optimised; a sensible choice for the bias factor already helps significantly to speed up the convergence of estimator of observables involving strangeness production. By starting with a sensible choice, the safeguard presented above is able to finalise the optimisation. A work on the bias optimisation was carried out. However, the observables multiplicity and the possibilities of initial parameters being too large, a simple way to determine the optimal bias factor cannot be provided by the authors. The simplest way to evaluate the optimal bias factor is an interpolation taking into account the energy and the mass number of the target, given that it is safer to be below the optimal bias factor. Examples of sensible choices can be found in \autoref{results} for each figure plotted.

In \autoref{problem}, we show the results of the study on the limits of the variance reduction methods implemented in INCL. The limits discussed are problematic only for some \textit{extreme} (high projectile energy and high target mass) reactions where the choice of the bias factor was not sensible. However, this case illustrates perfectly the problem of the importance dispersion, which produces large uncertainties in terms of variance \textit{jumps} even when the statistics is increased.

Here, we studied the mass distribution of hyperremnants in $p~(10~GeV)+~^{208}Pb$ collisions. The bias factors used were $1$ (no variance reduction method used), $2$, $5$, $10$, $20$, and $50$. The number of events was the same ($10^7$) for all the calculations and the computational time was equivalent. The upper and lower panels in the left column shows the results obtained using the variance reduction scheme, with a bias factor $2$ and $10$ respectively, together with the result from the standard calculation. The increase of the statistics ($n\_Hyp$) and the normalised mean uncertainty as a function of the bias factor are displayed in the right column. The left panels clearly show the minimal variance (\textit{i.e.} the smallest uncertainties) is not achieved with the highest statistics in this case.

An interpolation (shown as a blue line) of the normalised uncertainties predicts an optimal bias factor of around $2.5$. This is illustrated by the significantly better description of the hyperremnants spectrum obtained using a bias factor $2$ compared to the calculation using a bias factor $10$. The large uncertainties obtained using a bias factor $10$ are purely linked to the dispersion of hyperremnant importances. It is interesting to note that the interpolation of the effective bias factor (blue line) in the top right panel deviates from the nominal bias factor (dashed brown line) around the optimal bias factor ($2.5$). The effective bias factor is reduced by the safeguard discussed before and the reduction starts to be significant slightly above the optimal bias factor. This demonstrates that the safeguard worked well in this case, though it is not perfect.

\begin{figure}[!t]
\centering
\includegraphics[width=\columnwidth]{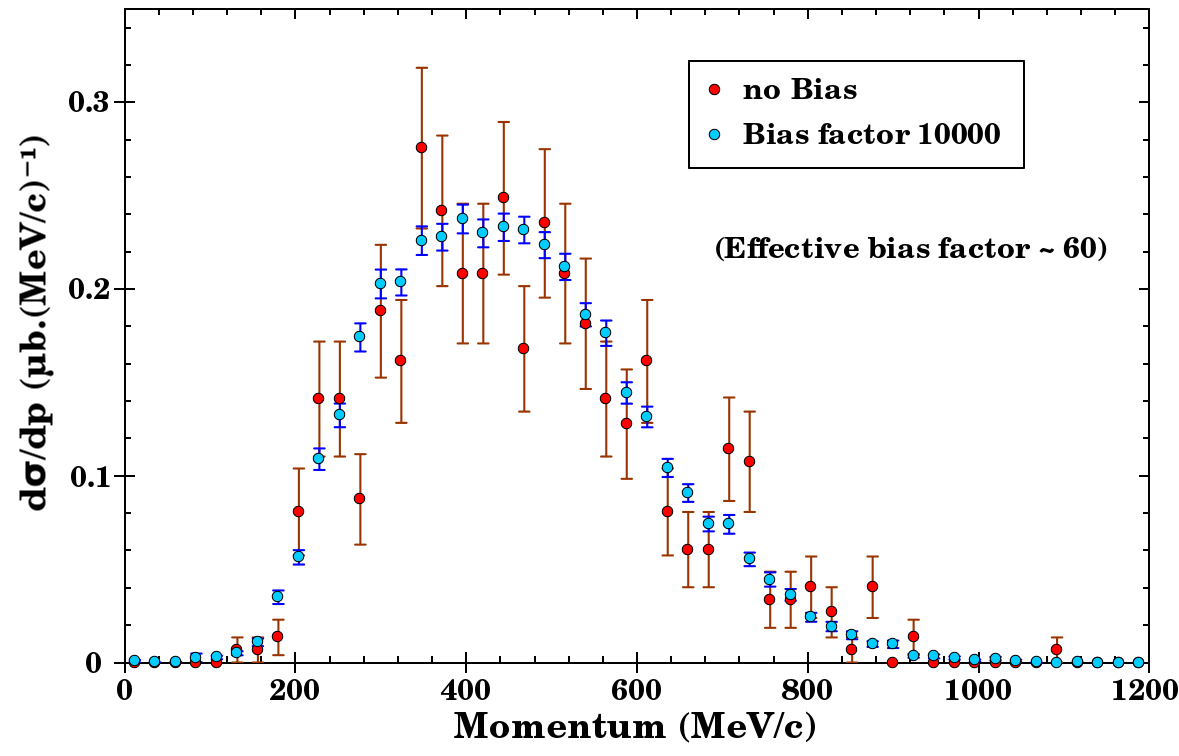}
\caption{\label{good} $K^+$ momentum in $p(1.6~GeV)+^{12}C$ with $10^7$ events. INCL using a bias factor $10^4$ (blue) is compared to INCL using no bias (red).}
\end{figure}

\begin{figure*}[t]
\begin{minipage}{1.95\columnwidth}
\begin{flushright}
\includegraphics[width=0.89\columnwidth]{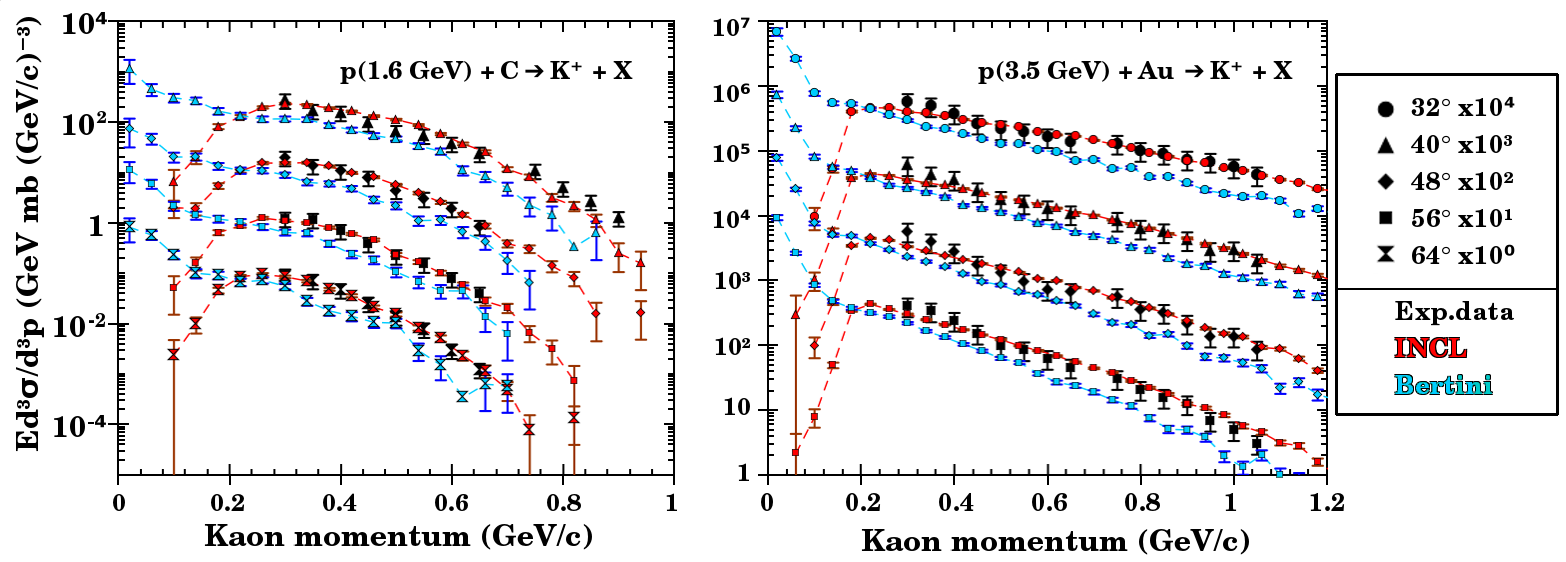}
\caption{\label{kaos_plot} $K^+$ invariant cross section for various angles in (left) $p(1.6~GeV)+C$ and (right) $p(3.5~GeV)+Au$ collisions. Experimental data \cite{kaos} (black symbols) are compared to INCL (red) and to the Bertini cascade model \cite{bertini} (blue). Bias factor used: $10$.}
\end{flushright}
\end{minipage}
\begin{minipage}{.05\columnwidth}
\end{minipage}
\end{figure*}

Generally, variance \textit{jumps} result in a global underestimation compared to the standard calculation and in some strongly overestimated bins. This can be dangerous if an underestimated bin is studied because the error bars for this bin are not significantly large but they can be far from the true value (see bottom left panel in \autoref{problem}). This is why it is crucial to minimise the importance dispersion. Remember, however, that every calculation converges to the same limit. Once again, the variance \textit{jumps} just make the convergence slower. The problem is that the variance, and by extension the error bars, might be underestimated because of the bad sampling.

\autoref{good} shows a reaction in which the number of nucleons in the target is too low to produce variance \textit{jumps} thanks to the safeguard regardless of the bias factor. The number of target nucleons being low, this leads to short intranuclear cascades with a low number of vertices. Additionally, the cross section ratio of vertices is limited by the safeguard. This results in strong constraints for the importances of final particles.

In such cases associated to a low energy, the safeguard matches automatically the effective bias factor to the optimal bias factor when the bias factor has been chosen too high. This is well illustrated in \autoref{good}. Although the nominal bias factor ($10^4$) is clearly above the optimal bias factor, which is $60$, the spectrum obtained using the VRS exhibits no variance \textit{jump}.

\section{Results}
\label{results}

In this section, results of INCL++6 calculations are compared to experimental data and to other models in order to facilitate the interpretation and the analysis. The different strengths and weaknesses observed in INCL++6 are discussed. Hypernucleus production is not presented here because it depends strongly on the de-excitation stage of the spallation reaction. A following paper will be dedicated to the strange degree of freedom in the de-excitation model and the hypernucleus production will be discussed in this publication. Here, we focus on the intra-nuclear cascade and single particle production.

The studied particles are the charged kaons ($K^+$ and $K^-$), the $\Lambda$, and the neutral kaons. The $K^+$ was clearly the most studied particle in the past. Experimental data exist near the threshold, and even in the sub-threshold region, up to high energies ($\sim14$~GeV). Additionally, various targets were studied and the emitted $K^+$ were observed at different angles. Moreover, the $K^+$ is the only particle together with the $K^0$ in our energy range that carries a positive hypercharge. Therefore, every strangeness production results in kaon production. Because the production modes of the two kaons are similar, studying the $K^+$ also gives an estimate of the reliability of the total strangeness production in INCL++6. The $K^-$ was less intensively studied because of the lower production rate (about two orders of magnitude compared to the $K^+$). However, double differential experimental data can be found in the literature. The $\Lambda$ and the neutral kaons having no electric charge, their detection is more complex and less experimental data are available. However, their analysis can help to understand the different processes in competition in the strangeness production.

The bias factors (see \autoref{biasing}) used to obtain the INCL++6 results are given in the captions of the corresponding figures.

\subsection{KaoS}
\label{kaos_sec}

The KaoS \cite{kaos} (Kaon Spectrometer) experiment was performed at the heavy-ion synchrotron SIS at GSI in Darmstadt. The KaoS collaboration measured the $K^+$ and $K^-$ production in $p+C$ and $p+Au$ collisions at $1.6$, $2.5$, and $3.5$~GeV proton beam kinetic energies. The kaon momentum was measured from $p_{lab}=0.3$ to $1.1$~GeV/c.

Most of the data measured for the $K^+$ production are well described by INCL. \hyperref[kaos_plot]{Fig.~\ref*{kaos_plot}} shows the $K^+$ production invariant cross section in two configurations measured by the KaoS collaboration in comparison to INCL and to the Bertini cascade model \cite{bertini}. Either for the left panel with $K^+$ production near threshold on a light nucleus (carbon) or for the right panel with $K^+$ production at higher energy on a heavy nucleus (gold), the results of INCL match very well the experimental data. The comparison with the KaoS data indicates a reliable total strangeness production cross section for a relatively large range of nuclei, energies, and angles. It can be seen that the Bertini cascade model gives a shape similar to INCL in the momentum range of the experimental data but underestimates them by roughly $40$\%. This difference can be explained by the $\Delta$-induced strangeness production, which is not included in the Bertini model. \autoref{kaos_plot} depicts for the low momenta region huge differences between the predictions from the INCL and Bertini models. This is due to the different values used for the $K^+$ potential. In our approach, the $K^+$ repulsive potential in INCL reduces drastically the invariant cross section at low momenta. Experimental data at lower momenta would help to test the $K^+$ potential.

\begin{figure}[t]
\begin{center}
\includegraphics[width=0.96\columnwidth ,keepaspectratio]{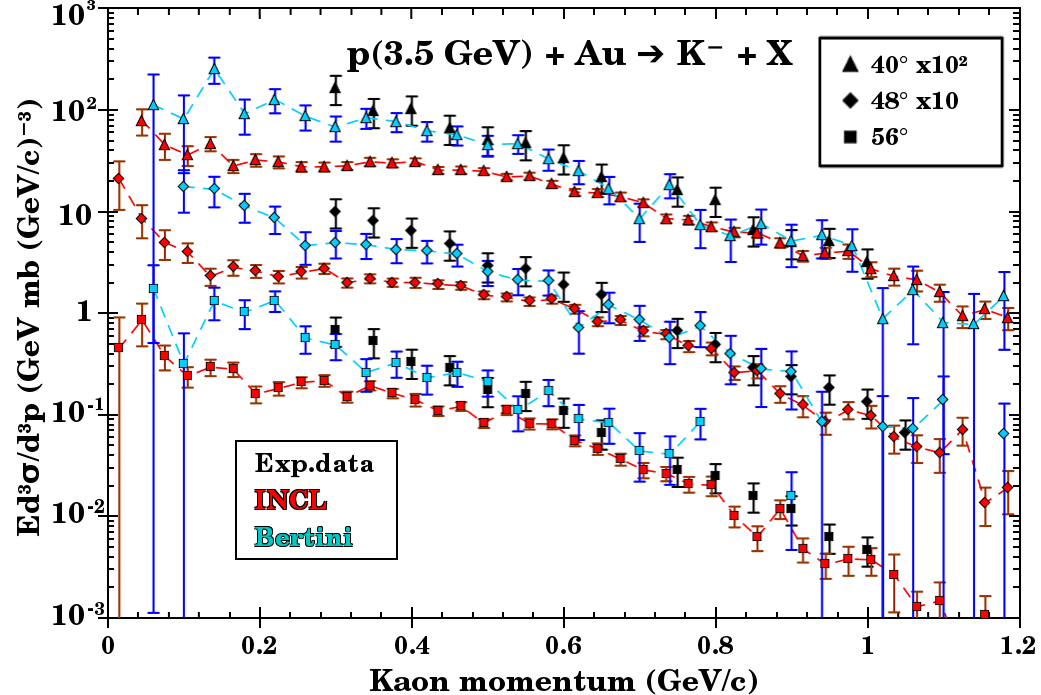}
\end{center}
\caption{\label{kaos_k-} $K^-$ invariant cross section at $40^\circ$, $48^\circ$, and $56^\circ$ in $p(3.5~GeV)+Au$ collisions. Experimental data \cite{kaos} (black) are compared to INCL (red) and to the Bertini cascade model \cite{bertini} (blue). Bias factor used: $10$.}
\end{figure}

\begin{figure*}[t]
\begin{minipage}{2\columnwidth}
\begin{center}
\includegraphics[width=0.8\columnwidth ,keepaspectratio]{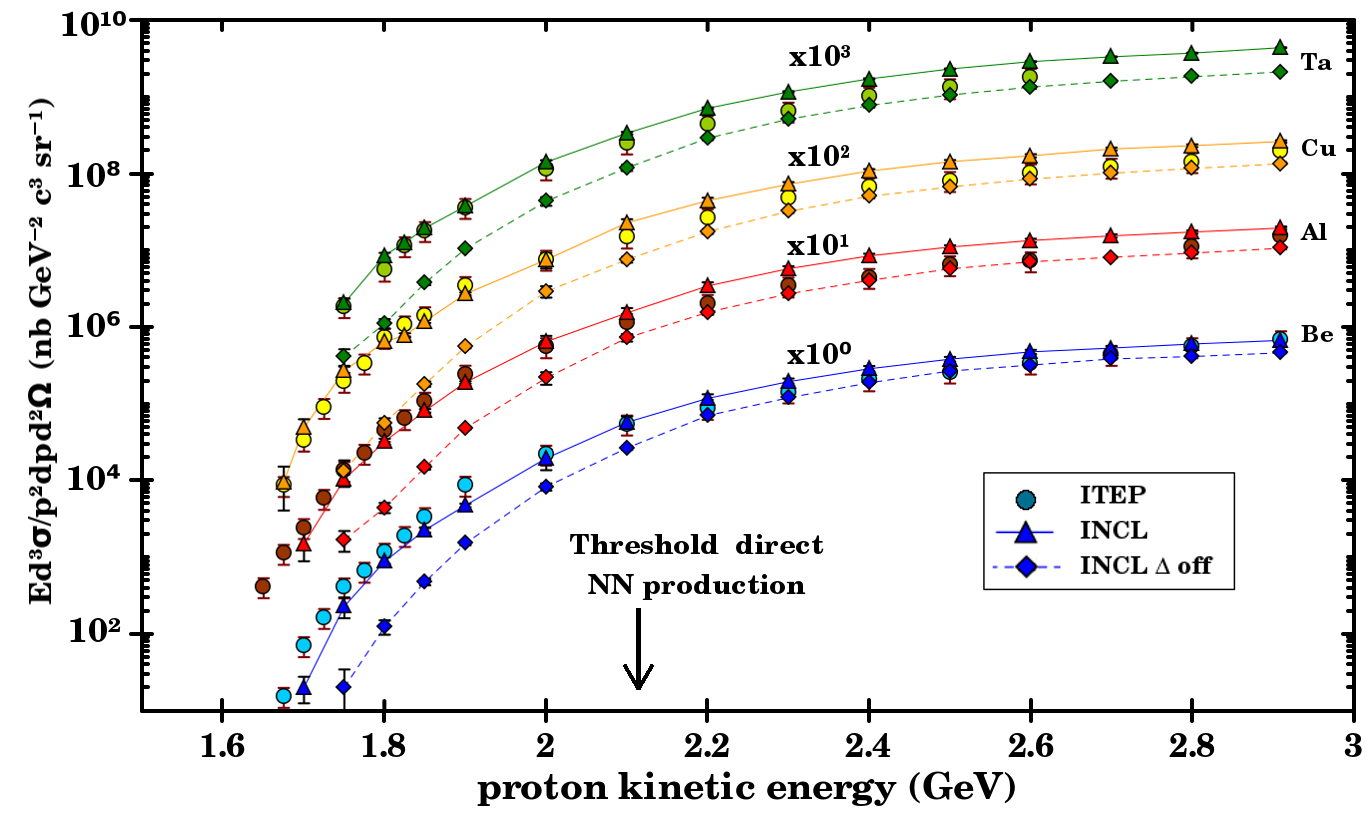}
\caption{\label{itep_plot} $K^+$ invariant cross section in $p+A\rightarrow K^++X$ reactions for kaons emitted with a momentum of $1.280 \pm 0.014~GeV/c$ at $\theta = 10.5^\circ $. The experimental data from \cite{itep} (circles) are compared to INCL with (triangles) and without $\Delta$-induced strangeness production (squares). Bias factor used: $50$ for INCL without $\Delta$-induced strangeness production for $T_p < 2~GeV$, $20$ otherwise.}
\end{center}
\end{minipage}
\end{figure*}

\autoref{kaos_k-} shows the invariant cross section for $K^-$ production for the reaction proton on $Au$ at $3.5$~GeV. It can be seen that the data are well described by INCL for momenta above $0.5$~GeV/c. However, below this energy, the cross section is clearly underestimated and INCL does not reproduce the shape of the experimental data. In ref.~\cite{kaos}, the authors concluded that the $K^-$ production is mostly due to the $NY\rightarrow~K^-NN$ reaction and to a lesser extent to the $\pi Y \rightarrow~K^-N$ reaction. Both reactions are not considered in INCL because of the lack of experimental data. This is probably the explanation. The low energy production being dominated by secondary reactions and the high energy production by primary reactions, our result seems consistent with a predominance of antikaon production through strangeness exchange ($NY\rightarrow~K^-NN$ and $\pi Y \rightarrow~K^-N$) for low momenta and through direct production ($NN\rightarrow NNK\overline{K}$) for high momenta. The Bertini cascade model, which includes the $NY\rightarrow~K^-NN$ (but not the $\pi Y \rightarrow~K^-N$) reaction, shows a significantly better description over the entire momentum range covered by the experimental data. This would be in favour of adding the extra strangeness exchange reactions even when considering the bad quality of experimental data.

\subsection{ITEP}
\label{ITEP_sec}

The study carried out at the Institute of Theoretical and Experimental Physics (ITEP) accelerator in Russia \cite{itep} measured the $K^+$ production in proton-nucleus collisions. The nuclei studied were $Be$, $Al$, $Cu$, and $Ta$. This choice covers a mass range from $A=9$ to $181$ ($Z=4$ to $73$). The projectile energy range from $1.65$ up to $2.91$~GeV. The experiment measured the production of $K^+$ with a momentum $p=1.280\pm 0.014$~GeV/c and with an emission angle $\theta = 10.5^\circ$. This very specific phase space constraint was used to drastically reduce the contribution of the $K^+$ production in $\pi N \rightarrow Y K$ secondary reactions (see ref.~\cite{itep} for details). This simplified their analysis by considering only the $N N \rightarrow N Y K$ primary reactions.

In our analysis, we tested the implemented $\Delta$-induced strangeness production, which is not based on experimental data, but the corresponding cross sections are based on theoretical calculations from Tsushima et al.~\cite{tsushima} (see ref.~\cite{bibi} for more details). In \autoref{itep_plot} we compare INCL calculations with and without $\Delta$-induced strangeness production to experimental data \cite{itep}. This allows to study the impact of $\Delta N \rightarrow N Y K$ secondary reactions compared to $N N \rightarrow N Y K$ primary reactions; the $\pi N \rightarrow Y K$ secondary reactions being naturally suppressed by the phase space constraint. The Bertini model, which has no variance reduction method, is not plotted in  \autoref{itep_plot} because of the unreasonable computing time needed to get comparable uncertainties. The threshold for the direct production of $K^+$ with a momentum $p=1.280\pm 0.014$~GeV/c is $T=2.115$~GeV in nucleon-nucleon collisions. The sub-threshold production is allowed thanks to effects of structure like the Fermi momentum, and also to secondary reactions (\textit{e.g.}, $\Delta N$) as explained below.

\autoref{itep_plot} shows that the standard INCL calculations (with $\Delta$-induced strangeness production) reproduce relatively well the shape and the absolute values of the experimental data, especially for energies below $2.1$~GeV, which correspond to the sub-threshold production. However, the standard INCL model overestimates the $2.1-2.9$~GeV energy region by around $50\%$. In comparison, INCL without $\Delta$-induced strangeness production underestimates the experimental data over the entire energy range. Above $T=2~GeV$, the underestimation is around $20\%$. This underestimation reaches an order of magnitude for the lowest momenta. This demonstrates the crucial role of $\Delta$'s in the strangeness production. Going deeper in the analysis, the overestimation in the $2.1-2.9$~GeV energy range indicates that the $\Delta$-induced kaon production in this region is likely overestimated. This seems consistent with observations made in ref.~\cite{tsushima} where the authors observed an overestimation of the cross sections for these reactions with center-of-mass energies $200$~MeV above the threshold.

The assumption that cross sections are very well described near the threshold but are overestimated at higher energy could be due to the negligence of hyperonic resonances in the model used in ref.~\cite{tsushima}. This choice was made because of the low confidence level associated to these resonances. However, they could play a significant role, notably at high energies.

A precise evaluation of $\Delta$-induced strangeness production cross sections with INCL is difficult to realise because of the complexity of the spallation process. However, the impossibility to measure the $\Delta$-induced strangeness production cross sections and the limitations of the theoretical calculations make simulation models like INCL suitable candidates to estimate and/or test such cross sections.

\subsection{ANKE}
\label{ANKE_sec}

\begin{figure}
\begin{center}
\includegraphics[width=\columnwidth ,keepaspectratio]{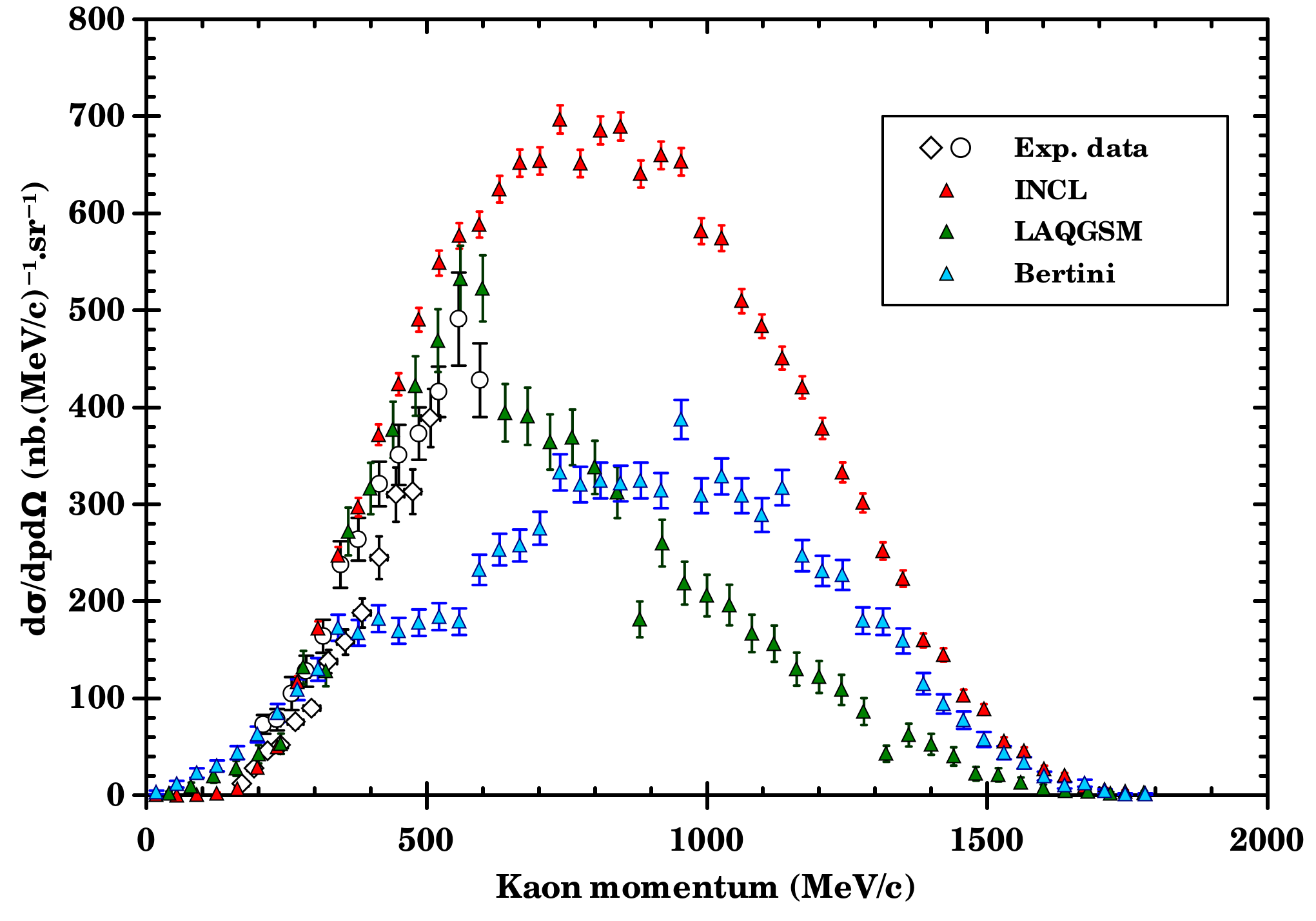}
\end{center}
\caption{\label{anke_plot} $K^+$ momentum spectrum in $p(2.3~GeV)+^{12}C$ collisions within the angular acceptance of the ANKE experiment. Two sets of experimental data \cite{anke} (circles and squares) are compared to INCL (red), LAQGSM \cite{laqgsm} (green), and the Bertini cascade model \cite{bertini} (blue). Bias factor used: $10$.}
\end{figure}

The ANKE experiment \cite{anke} investigated the production of $K^+$ in the forward direction in $p+A$ collisions with proton kinetic energies between $1$ and $2.3$~GeV. The targets were $A =\ ^{2}\!H, ^{12}\!C, Cu,$ $Ag$, and $^{197}\!Au$. The experiment took place at the COoler SYnchrotron COSY-Jülich in Germany. The angular acceptance was $\pm 12^\circ$ horizontally and $\pm 7^\circ$ vertically.

In \autoref{anke_plot}, calculations from LAQGSM \cite{laqgsm}, Bertini \cite{bertini}, and INCL are compared to ANKE experimental data. It can be seen that the three models fit relatively well the experimental data at low momenta (below $300$~MeV/c) with INCL being slightly closer to the experimental data than the other two. At higher momenta, INCL and LAQGSM still reproduces the data while Bertini underestimates them. At higher energies where no experimental data exist, every model gives a different shape. Whereas LAQGSM decreases quickly, INCL continues to increase and Bertini predicts a bump. The maxima are around $600$~MeV/c, $800$~MeV/c, and $900$~MeV/c for LAQGSM, INCL, and Bertini, respectively. An extrapolation from \autoref{itep_plot} suggests that INCL overestimates the production cross section for $K^+$ with $p_{K^+}=1.280$~GeV/c in forward direction by roughly $30\%$. This would be compatible with the Bertini's value at $p_{K^+}=1.280$~GeV/c. Again, the INCL $\Delta$-induced kaon production is probably overestimated.

\subsection{LBL}
\label{LBL_sec}

\begin{figure}[t]
\centering
\includegraphics[width=\columnwidth]{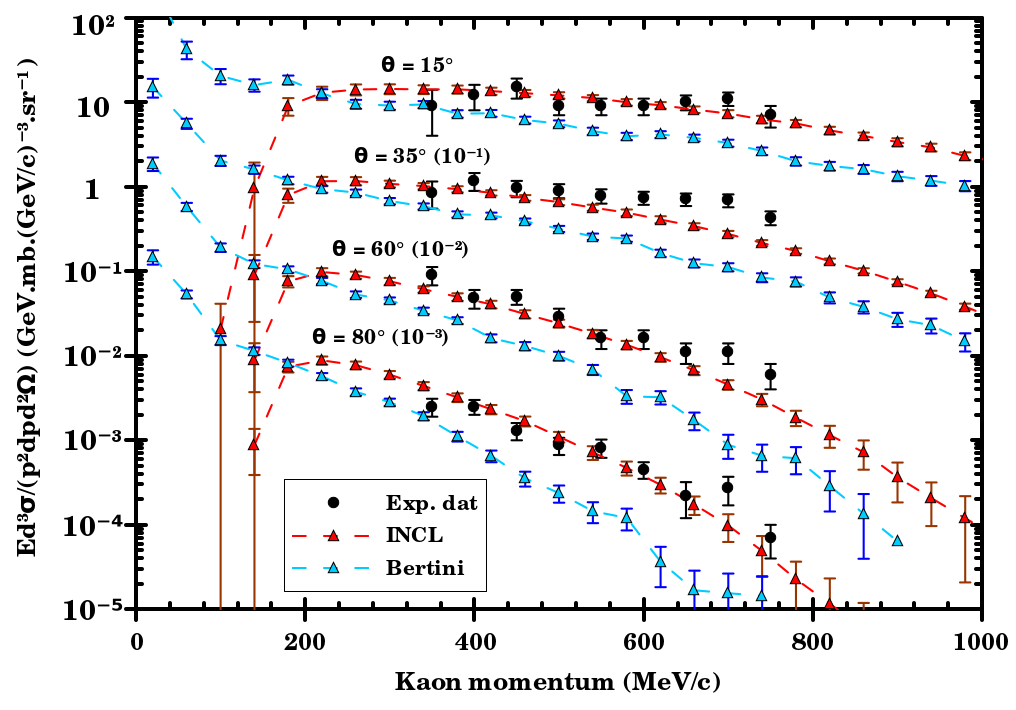}
\includegraphics[width=\columnwidth]{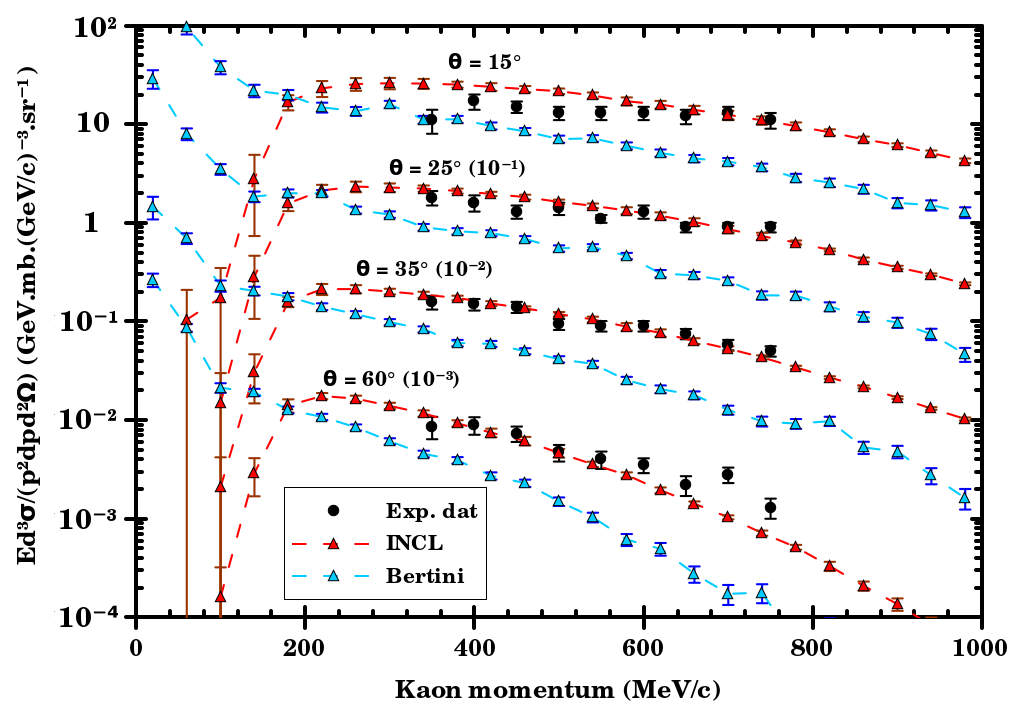}
\caption{\label{lbl_plot} $K^+$ invariant cross section at $2.1$~GeV/Nucleon for the reactions $p+^{208}Pb$ (upper panel) and $^2H+^{208}Pb$ (lower panel). The experimental data \cite{lbl} (black dots) measured at the LBL are compared to the Bertini cascade model \cite{bertini} (blue) and INCL (red). Bias factor used: $10$.}
\end{figure}

The experiment carried out at the Lawrence Berkeley Laboratory (LBL) \cite{lbl} studied the inclusive $K^+$ production using projectiles with $T=2.1~GeV/nucleon$. Several projectile types and targets were tested but a direct comparison between measured and modelled data is only possible for the reactions $p+^{208}Pb$ and $^2H+^{208}Pb$. 

In order to estimate the inclusive $K^+$ production cross section, the collaboration measured $K^+$ spectra at four different angles: $\theta = 15^\circ$, $35^\circ$, $60^\circ$, and $80^\circ$ for the $p+^{208}Pb$ reaction and $\theta = 15^\circ$, $25^\circ$, $35^\circ$, and $60^\circ$ for the $^2H+^{208}Pb$ reaction. Spectra were measured for kaon momenta from $350$ to $750$~MeV/c.

The two panels in \autoref{lbl_plot} shows that there is good agreement between INCL predictions and experimental data both for proton and deuteron induced reactions. It must be mentioned this experiment is very similar to the one carried out by the KaoS collaboration. Though, the angles studied, the energies, the projectiles, and the targets are slightly different but the conclusions are the same. Again, the good agreement not only in the absolute values but also in the energy dependence of the data demonstrates the reliability of the handling of strange particles in the new INCL version. This is a first validation that INCL can handle light clusters as projectile.

\subsection{Hades}
\label{hades_sec}

The HADES (High Acceptance DiElectron Spectrometer) collaboration studied $\Lambda$ and $K^0_s$ productions in $p+p$ and $p+Nb$ collisions at $3.5$~GeV \cite{hades_l,hades_k}. The Experiment took place at GSI in Germany. The $\Lambda$ particles were measured in the $[0.1,1.3]$ rapidity ($y$) range in the laboratory frame and the $K^0_s$ particles were measured in the $[-0.85,0.55]$ rapidity range in the nucleon-nucleon center of mass. In this second case, the rapidity of the laboratory is $y_{lab}=-1.118$.

The hypernucleus production is strongly correlated to the $\Lambda$ production since most of the observed hypernuclei involve one or more $\Lambda$'s. Therefore, the $\Lambda$ production must be tested before studying the hypernucleus production, which implies more complex processes. 

\begin{figure}[t]
\centering
\includegraphics[width=\columnwidth]{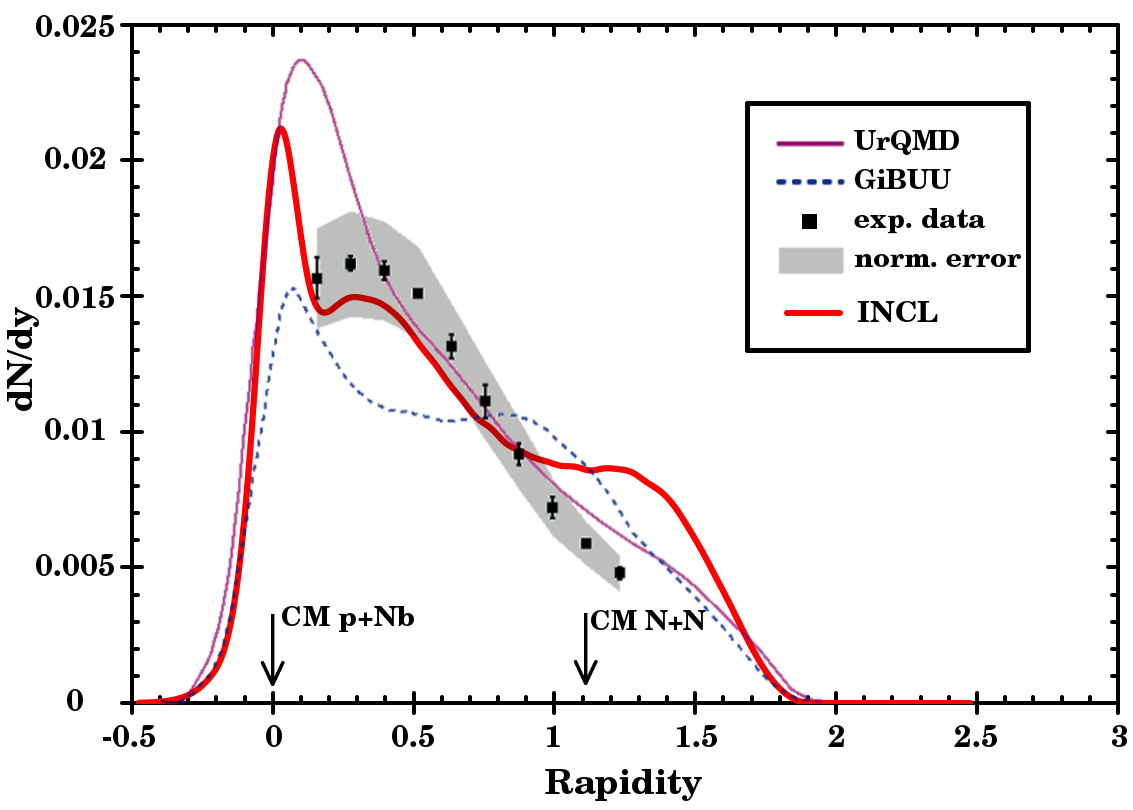}
\caption{\label{hades_plot_lambda} $\Lambda$ production yield in $p(3.5~GeV)+Nb$ collisions as a function of rapidity. The HADES experimental data (black square) are compared to GiBUU (blue dashed line), UrQMD (purple line), and INCL (red line) model predictions. The original plot can be found in \cite{hades_l}. Bias factor used: $10$.}
\end{figure}

In \autoref{hades_plot_lambda} we compare the $\Lambda$ production yield as function of rapidity calculated using INCL to predictions from UrQMD \cite{UrQMD} and GiBUU \cite{gibuu} and to experimental data \cite{hades_l}. The original plot is taken from \cite{hades_l}. The Bertini cascade model does not handle $\Sigma^0$'s decay, which plays a role in the $\Lambda$ production. Therefore, results from the Bertini code are not plotted here.

\begin{figure}[t]
\includegraphics[width=\columnwidth]{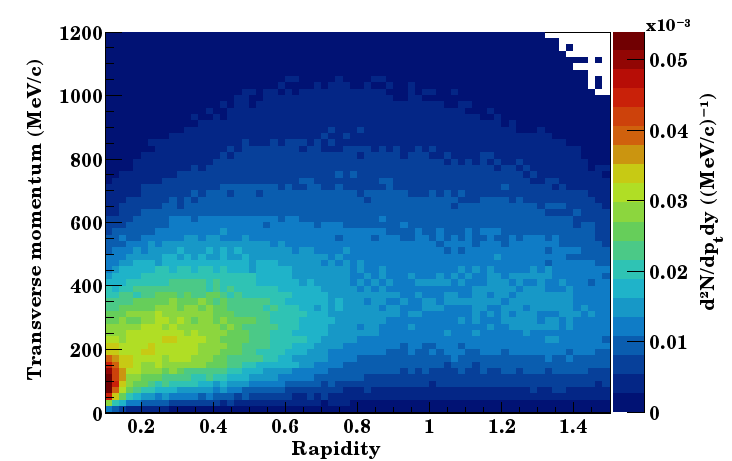}\\
\includegraphics[width=\columnwidth]{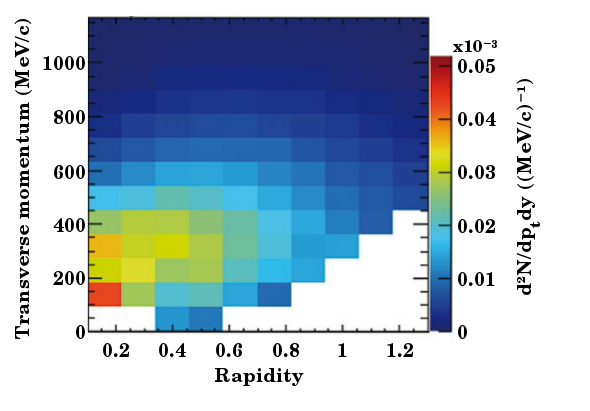}
\caption{\label{hades_plot_lambda_2} $\Lambda$ rapidity versus transverse momentum yield in $p(3.5~GeV)+Nb$ collisions. Top: INCL predictions. Bottom: Experimental data from \cite{hades_l} after correction of efficiency.}
\end{figure}

\begin{figure}
\centering
\includegraphics[width=\columnwidth]{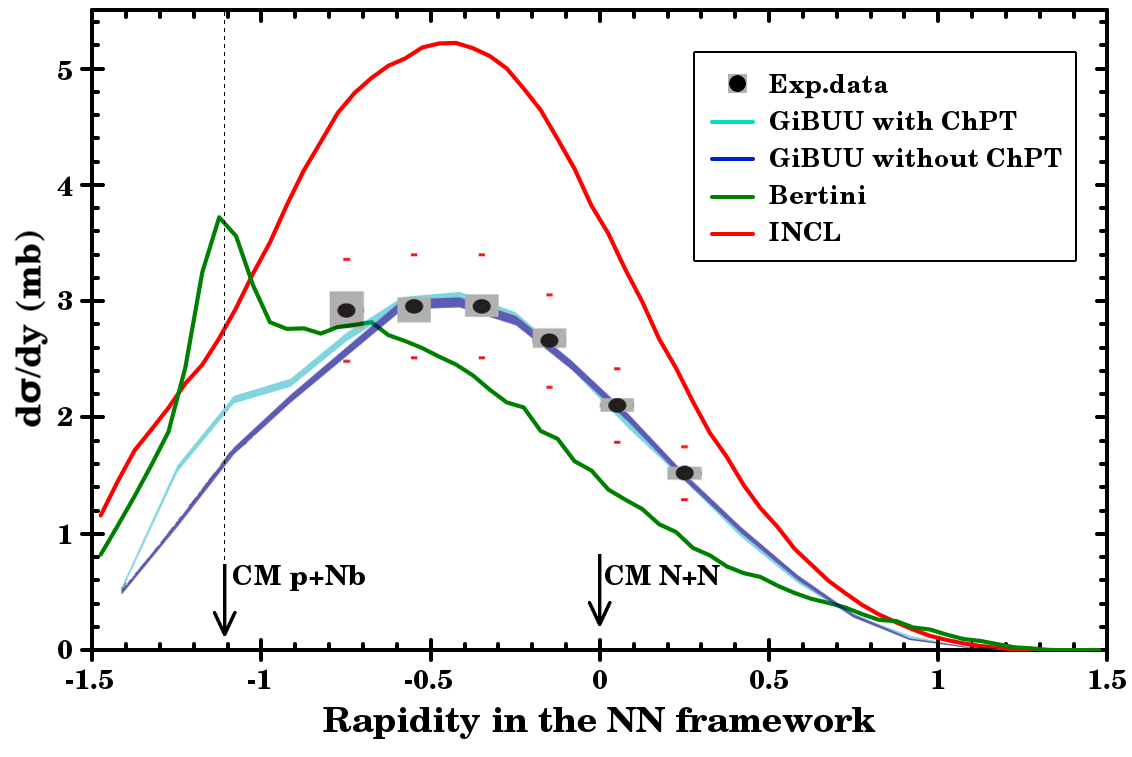}
\caption{\label{hades_plot_k0} $K^0_s$ production cross section in $p(3.5~GeV)+Nb$ collisions in function of the rapidity in the nucleon-nucleon center of mass. HADES experimental data \cite{hades_k} (black circles) are compared to GiBUU with (cyan) and without (blue) a chiral potential, the Bertini cascade model (green) and INCL (red). The little red dots represent the experimental systematics linked to the normalisation. The original plot comes from ref.~\cite{hades_k}. Bias factor used: $10$.}
\end{figure}

\autoref{hades_plot_lambda} depicts a rather good agreement between INCL and the experimental data if taking into account systematic errors except for the rapidities higher than $0.9$. The bump observed in the INCL predictions around $y=1.3$ can also be seen with the GiBUU model, but at lower rapidity. The UrQMD model does not exhibit such a behaviour and is close to the experimental data, but it misses the first two data points at the lowest rapidities.

Aiming at understanding the INCL bump, we show in \autoref{hades_plot_lambda_2} the transverse momentum versus the rapidity of $\Lambda$ particles. It can be seen that the top (INCL) and bottom (HADES data) panels match well. For the problematic rapidities (above $y=0.9$) one can find an overabundance of $\Lambda$ particles of INCL compared to HADES in the rapidity range $[1,1.4]$ and the transverse momentum range $[200,500]$~MeV/c. This corresponds to emission angles between $5$ and $19$ degrees when the experiment acceptance is between $18$ to $85$ degrees.

The acceptance of the HADES experiment being limited, data for the unmeasured regions visible in the lower panel of \autoref{hades_plot_lambda_2} have been estimated by extrapolating the transverse momentum spectra using Maxwell-Boltzmann distributions. The thus determined data are in disagreement with the INCL prediction. Consequently, the bump around $y=1.3$ predicted by INCL is in a phase-space not measured experimentally. Therefore, there is no strict contradiction between experimental data and INCL model predictions

In the region around $y=0$, every model predicts a peak. This peak is high and broad using UrQMD, narrow and smaller using GiBUU, and high and narrow using INCL. A comparison between INCL with and without $\Sigma^0$ in-flight decay showed the peak in INCL predictions is entirely due to the $\Lambda$ production inside the nucleus. Additionally, studying the origin of the $\Lambda$'s indicated that this peak is the result of the hyperremnant de-excitation for INCL.

The second particle measured by the HADES collaboration is the $K^0_s$. In \autoref{hades_plot_k0} the experimental data \cite{hades_k} are compared to GiBUU calculations with and without a chiral potential, to the Bertini cascade model, and to INCL. It can be seen that the best description of the experimental data are obtained using GiBUU. However, the version of GiBUU used here is a modified version in which the $K^0$ production has been artificially reduced to fit the $p+p$ HADES experimental data (see ref.~\cite{hades_k} for details). Additionally, this GiBUU version is not the same as the one used in ref.~\cite{hades_l}, although the reaction studied is the same. This makes their results difficult to interpret. INCL gives a good shape but overestimates the experimental data by roughly $65\%$. Similarly to the case for the $K^+$ production, the $\Delta$-induced reactions could be an explanation for the overestimation but some part of this overestimation might also be due to the normalisation. The total reaction cross section $\sigma^{p+Nb}_{tot}$ used by HADES is $848 \pm 127~mb$, while INCL calculates a value of $1048~mb$. A measurement of the total reaction cross section for the same system at a lower energy ($1.2$~GeV instead of $3.5$~GeV) gave $1063 \pm 40~mb$~\cite{herbach}, which is consistent with the INCL value. The Bertini cascade model does not reproduce the energy dependence of the experimental data, but the predicted absolute yield of $K^0_s$ corresponds to the experimental data. The bump around the laboratory rapidity ($y_{lab} = -1.118$) can also be seen in GiBUU predictions with the chiral potential. This is likely due to the attractive potential used by these models for the $K^0$, whereas INCL uses a repulsive potential.

\subsection{FOPI}
\label{fopi_sec}

The FOPI collaboration \cite{fopi} measured the in-medium neutral kaon inclusive cross sections in $\pi^-$-induced reactions on various targets: $C$, $Al$, $Cu$, $Sn$, and $Pb$. The pion beam had a kinetic energy $T \simeq1.02$~GeV $(p_\pi=1150~MeV/c)$. The experiment took place at the heavy-ion synchrotron SIS at GSI. The geometrical acceptance of the detector was restricted to the polar angles in the range $25^\circ < \theta < 150^\circ$. An a posteriori correction was applied by the FOPI collaboration to obtain the total inclusive cross sections.

\begin{figure}[t]
\centering
\includegraphics[width=\columnwidth]{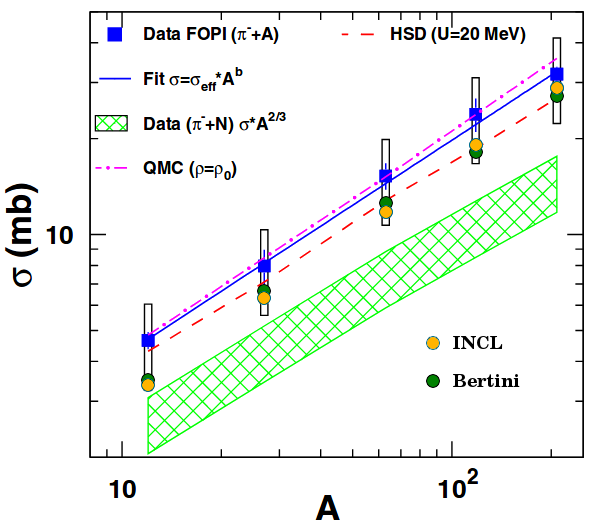}
\caption{\label{fopi_plot} Neutral kaon production cross section in the collision $\pi^-(1020~MeV)+A$ $(p_\pi=1150~MeV/c)$. The FOPI experimental data \cite{fopi} (blue squares) are compared to QMC \cite{qmc} (pink dash-dotted line), HSD \citep{hsd1,hsd2} (red dashed line), Bertini \cite{bertini} (green circles), and INCL (orange circles) models. The meaning of the hatched area is explained in the text. The original figure comes from ref.~\cite{fopi}. Bias factor used: $10$.}
\end{figure}

\autoref{fopi_plot} depicts the production of neutral kaons measured by the FOPI collaboration. In the original paper \cite{fopi}, the authors compared their experimental data with two models: the quark-meson coupling model (QMC) \cite{qmc} and the hadron-string-dynamics model (HSD) \citep{hsd1,hsd2}. By fitting the data, the FOPI collaboration found that the inclusive cross sections can be described by a $A$-dependent cross section of the form $\sigma(A)=\sigma_{eff} \! \times \! \ A^{b}$, with $b = 0.67 \pm 0.03$. This would indicate that $K^0$ production is dominated by peripheral collision. Therefore, they developed the function:
\begin{equation}
\label{supo}
\sigma(A) = \sigma(\pi^-(1150~MeV/c) + N \rightarrow K^0 + X) \times A^{2/3},
\end{equation}
with $N$ a target nucleon. The nucleon cross section was obtained by summing all $\sigma(\pi^-+N \rightarrow K^0 + Y)$ processes weighted with the relative proton and neutron numbers of the target nucleus. The thus developed function is represented as a hatched band in \autoref{fopi_plot}. The bandwidth corresponds to an uncertainty of $20\%$. It appears that the experimental data do not enter in the region defined by this band, which indicates that the simple dependence expressed in \autoref{supo} cannot explain the observed cross sections. Therefore, there must be an additional process.

The INCL calculations slightly underestimate the experimental data but they are within the experimental systematic uncertainties (rectangular bars). The inclusive cross sections predicted by INCL is proportionnal to $A^{3/4}$, which promote a different interpretation on the $K^0$ production processes with larger contributions of secondary reactions and a deeper strangeness production. The Bertini cascade model \cite{bertini} shows a result similar to INCL but with a slightly lower slope. Both INCL and Bertini cross sections are consistent with \autoref{supo} using a dependence of $A^{3/4}$ instead of the proposed $A^{2/3}$.

\subsection{E-802}
\label{e802_sec}

\begin{figure}[!t]
\centering
\includegraphics[width=\columnwidth]{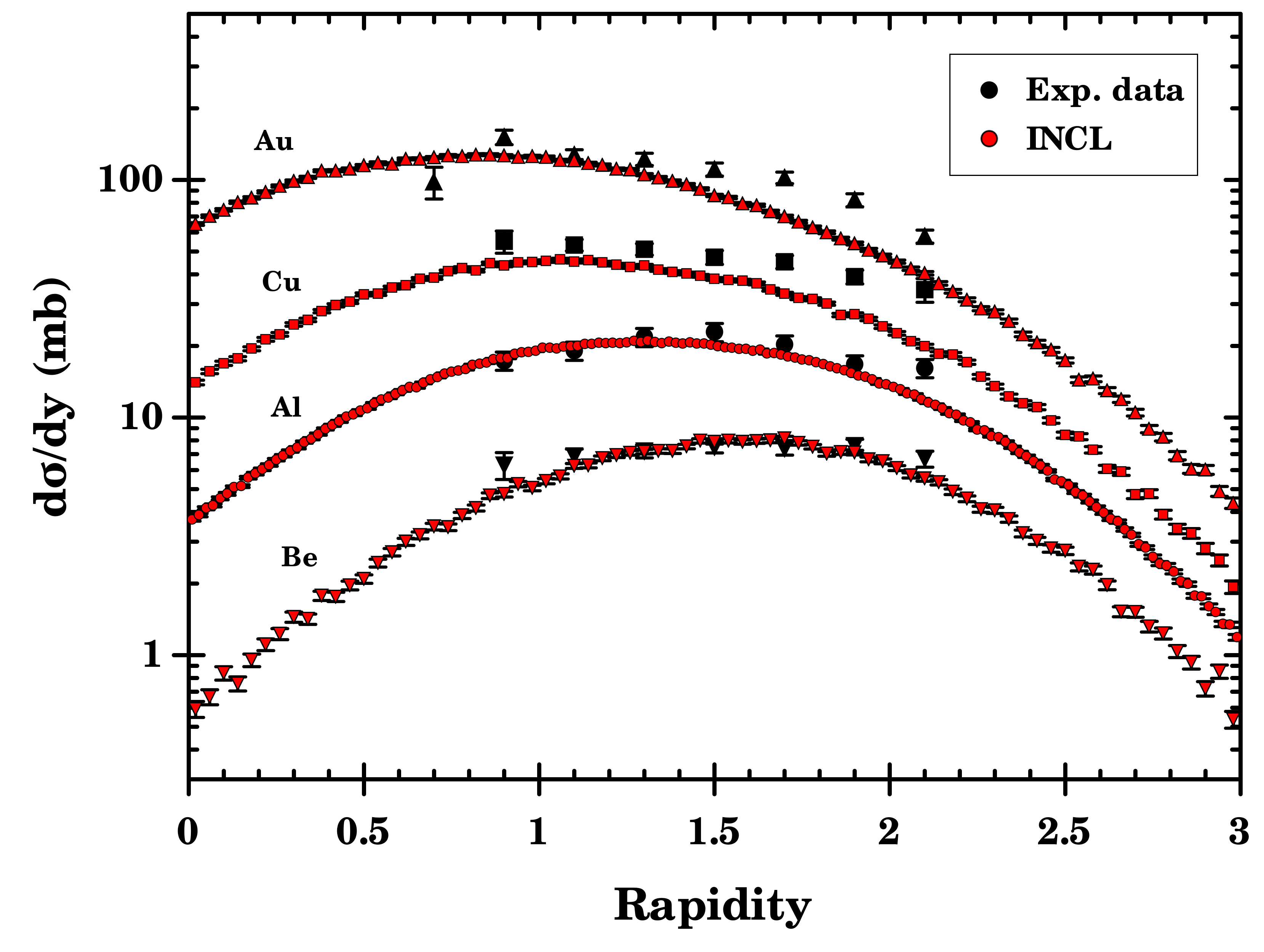}
\caption{\label{e802_plot} $K^+$ production cross sections in $p(14.6$~GeV/c$)+A$ collisions as a function of rapidity. The experimental data \cite{e802} (black) are compared to INCL predictions (red). No variance reduction method was used.}
\end{figure}

The INCL model has been extended to work up to incident energies of about $\sim15$~GeV. However, the experiment studying strangeness production focus below $3.5$~GeV. A notable exception is the experiment realised by the E-802 collaboration \cite{e802} at the Brookhaven National Laboratory (US) using a proton beam momentum of $14.6$~GeV/c $(T_p \simeq13.7~GeV)$. This experiment measured various particle production cross sections in proton-nucleus reactions and, notably, is also measured the $K^+$ production cross sections. Therefore, it provides a good opportunity to test INCL strangeness physics for the highest energies. The nuclei studied by the E-802 collaboration were $Be$, $Al$, $Cu$, and $Au$.

\hyperref[e802_plot]{Fig.~\ref*{e802_plot}} shows a comparison of INCL predictions with E-802 experimental data. The Bertini model predictions are not plotted because the upper energy limit for this model is about $10$~GeV. Considering that the experimental data are close to the upper limit of INCL, the shape as well as the absolute values are in excellent agreement. However, a slight underestimation at high rapidities can be observed for $Cu$ and $Au$. The observed deviations are difficult to analyse because they are only for two target elements and the discrepancies differ: a bump for $Au$ and a linear deviation for $Cu$.

\begin{figure}
\centering
\includegraphics[width=\columnwidth]{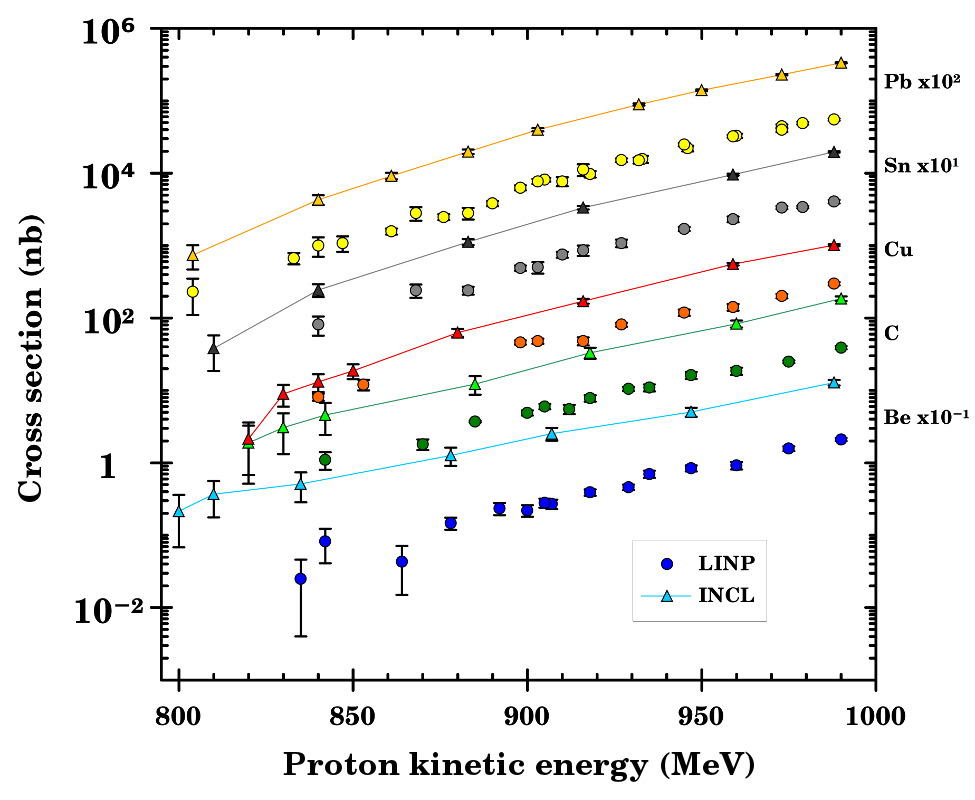}
\caption{\label{koptev_plot} Subthreshold $K^+$ production cross sections in $p+A$ collisions as a function of the initial proton kinetic energy. The experimental data \cite{linp} (circles) are compared to INCL predictions (triangles).}
\end{figure}

\subsection{LINP}
\label{linp_sec}

The experiment carried out at the Leningrad Institute of Nuclear Physics (LINP) \cite{linp} studied the sub-threshold $K^+$ production on various nuclei, from beryllium to uranium. A measurement of the total cross sections as a function of energy was performed in the $800-1000$~MeV energy range for $Be$, $C$, $Cu$, $Sn$, and $Pb$.

In \autoref{koptev_plot} INCL predictions are compared with the experimental data. The first observation is a clear overestimation of the experimental data. This overestimation is around a factor $6$ for $Be$ and about a factor $4$ for the other nuclei. The prediction of the slope is good, especially for $C$, $Sn$, and $Pb$. While the agreement between model predictions and experimental data is only fair, the major success of these calculations is the variance reduction (see \autoref{biasing}). The effective bias factors of these calculations are around $1000$ depending on the configuration with a maximum at $2000$. This allowed to obtain cross sections below the nano-barn scale using a relatively short calculation time (half a day for the entire set using parallelisation). Such a result cannot be obtained in a reasonable time without a variance reduction method.

Different explanations are possible to explain the overestimation. To understand the problem, it is important to remember that the beam energies in this experiment are far below the $K^+$ production threshold for $p+p$ collisions ($1.582$~GeV). Therefore, the strangeness production requires strong effects of structure, which are complex and not very well understood. Such effects include effects related to the momentum spectra of particles inside the nucleus and effects related to successive collisions. Additionally, $\Delta$ particles seem to play a crucial role in this experiment. An explanation for the overestimation could therefore be the treatment of the $\Delta$ particles in INCL. Another explanation could be the semi-classical description of the nucleus by INCL, which does not consider every quantum effect, which could play a significant role in sub-threshold processes. A following study has been carried out in order to better understand the discrepancies. The effect and limitations of the semi-classical approach cannot be tested but modifications related to the physics of $\Delta$ particles or to effects of structure showed a sensible reduction of the strangeness production. None of the modifications, however, could fully explain the discrepancies. Therefore, the overestimation is likely due to a combination of the three explanations proposed above. However, it is difficult to proof or reject this hypothesis considering the complexity of the processes involved.

\section{Conclusion and future}
\label{conc}

In this paper, a full description and a detailed analysis of the new version of INCL, INCL++6, were presented. First, the main new capabilities of INCL, which are the strangeness production and the variance reduction methods, were described. Then, the model predictions were compared to experimental data.

In this second part, different projectiles (proton, deuteron, and pion) were tested at various kinetic energies: from $1.6$ to $13.7$~GeV for protons, $1.02$~GeV for pions, and $2.1$~GeV/nucleon for deuteron. Various targets ranging from $^{9}Be$ to $^{208}Pb$ were considered and a wide range of angles was covered.

For most of the studied cases, there is a good agreement between experimental data and INCL predictions. Notably, the dependence of experimental data on either energy or target mass number is often well reproduced, which demonstrates that the strangeness physics is well implemented. However, slight discrepancies exist. In some cases experimental data are overestimated. Different explanations were proposed in order to understand these discrepancies, but the lack of experimental data prevents definitive conclusions. The dominant problem is likely in the $\Delta$-induced reactions. Though, implementing $\Delta$-induced reactions significantly improved the description of strangeness production, the cross sections used are suspected to be overestimated at high energies, which is consistent with the conclusions in ref.~\cite{tsushima}.

With the validation of INCL++6, the problem of the statistics and computational time arose. This problem has been solved with the implementation of a variance reduction scheme in INCL, which has been described in detail. The developed method increases artificially the strangeness production respecting some mathematical constraints, which results in a better statistics for observables linked to the strangeness physics, and therefore, to a reduction of the required calculation time.

In addition, our new version of INCL is now implemented in the transport code Geant4 \cite{geant4}. Thus, it can be used in the simulation of macroscopic systems. This also permits other collaborations to have a full access to the last version of our model. They can use it for the design of new experiments dedicated to the study of strange particles and hypernuclei in the foreseeable future, such as for the HypHI \cite{HypHI}, Panda \cite{Panda}, and CBM \cite{CBM} experiments at the FAIR facility.

In the future, a more detailed study of $\Delta$-induced strangeness production should be carried out to improve the corresponding cross section parametrisations. The $K^-$ production could be improved by the implementation of additional strangeness exchange channels. Concerning the variance reduction methods, the safeguard can be improved to reduce the number of edge cases exhibiting convergence issues, what would simplify its use.

\section*{Acknowledgements}

The authors would like to thank Davide Mancusi for useful discussions, Dennis Wright for Bertini model calculations, Joana Wirth from the HADES collaboration, Janus Weil from the GiBUU collaboration for useful discussions and Nikolai Mokhov for LAQGSM calculations.

This work has been supported by the Swiss National Science Foundation (SNF 200021-159562 and 200020-182447) and has been partially supported by the EU ENSAR2 FP7 project (Grant Agreement No. 654002).

\appendix

\section{Vertex cross section ratio}
\label{importance_ratio}

In this appendix some examples of cross section ratio calculations are discussed. This plans to explain the biasing steps of the variance reduction scheme.

\subsection{Example 1}
\label{ex1}

\begin{figure}[t]
\centering
\includegraphics[width=\columnwidth]{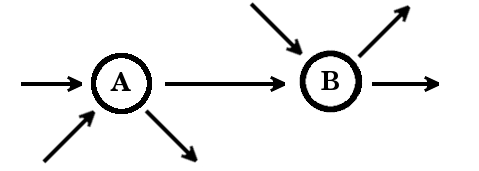}
\caption{\label{simple_case} Example of basic intranuclear cascade represented as a time ordered graph. Circles and arrows represent binary collisions and the particles propagating respectively.}
\end{figure}

In the first example (\autoref{simple_case}) we consider two successive binary collisions. The two non-strange particles at the origin of the vertex $A$ are considered without any history. Let us assume that the total cross section for the vertex $A$ is $20$~mb, with the reaction cross section for the production of strangeness is $0.1$~mb and therefore the non-strangeness production cross section is $19.9$~mb. Our goal is an augmentation of the total strangeness production by a factor $10$ (= bias factor).

According to \autoref{br}, to increase the probability of a particle production by a factor $10$ is to give it an importance $W=1/10$.

Since initial particles in $A$ have no history, the probability of strangeness production at this time of the cascade is directly proportional to the strangeness production cross section. Consequently, to increase the probability of strangeness production by a factor $10$, the corresponding cross section should be multiplied by the same factor $10$. Therefore, the modified strangeness production cross section is $1$~mb and, because the total cross section should be conserved, the modified non-strangeness production cross section is $19$~mb.


At this stage of the variance reduction scheme, the cascade is biased. Now, the vertex cross section ratio of the vertex $A$ should be determined. The vertex cross section ratio is the ratio of the event importance before versus after this vertex. This ratio is directly equal to the inverse of the cross section multiplying factor of the chosen channel. This means the vertex cross section ratio will be equal, in this case, to $0.1~mb/1~mb = 0.1$ if strangeness is produced and equal to $19.9~mb/19~mb \simeq1.047$ otherwise. Let us now assume that no strange particle is produced. At this stage, the probability of realisation for this cascade is decreased by a factor $1.047$ and the importances of outgoing particles are increased by $1.047$. In INCL, the cross section ratio is registered for future uses.

Let us assume that one of the particles produced in the vertex $A$ collides with another non-strange particle without history in the vertex $B$. Remember that we want to increase the probability of creating strange particles by a factor $10$. Therefore, the importance of a strange particle produced in the vertex $B$ should be $W=0.1$. The importance is the product of every cross section ratio on a path:

\begin{figure}
\centering
\includegraphics[width=\columnwidth]{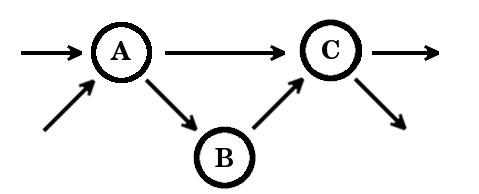}
\caption{\label{sec_case} Same as \autoref{simple_case} with three vertices and a branch recombination.}
\end{figure}

\begin{equation}
0.1 = W = CSR(A) \times CSR(B).
\end{equation}

Therefore, if a strange particle is produced, the cross section ratio of vertex $B$ should be $CSR(B)_S=1/(10 \times CSR(A)) \simeq 0.095$. This determines the modified strangeness production cross section of this vertex ($\sigma'_S = \sigma_S/CSR(B)_S$), the modified non-strangeness production cross section ($\sigma'_{NS} = \sigma_{tot} - \sigma'_S$), and the cross section ratio of vertex $B$ in case of non-strangeness production ($CSR(B)_{NS} = \sigma_{NS} / \sigma'_{NS}$).

In the general case, the cross section ratio of a vertex following a set $K$ of vertices and with strange particles in its final state should be:
\begin{equation}
\label{importance_calc}
CSR(X) = \frac{1}{\text{bias ratio} \times \prod_{I \in K} CSR(I)}.
\end{equation}

\subsection{Example 2}

The second example shown in \autoref{sec_case} illustrates the case of branch recombination. Continuing the case presented in \autoref{ex1}, a particle coming from vertex $B$, which is itself induced by a particle from vertex $A$, collides with a particle directly produced by vertex A.

The \autoref{importance_calc} gives $CSR(C)$ in case of strangeness production.
\begin{equation}
CSR(C) = \frac{1}{\text{bias ratio} \times CSR(A) \times CSR(B)}.
\end{equation}

We also know $CSR(B)$, which produced strangeness:
\begin{equation}
CSR(B) = \frac{1}{\text{bias ratio} \times CSR(A)}.
\end{equation}

Therefore $CSR(C) = 1$. Consequently, the cross sections are not modified because the probability to reach the initial state of vertex $C$ has already been increased by a factor equal to the bias ratio.

\subsection{Example 3}

\begin{figure}
\centering
\includegraphics[width=\columnwidth]{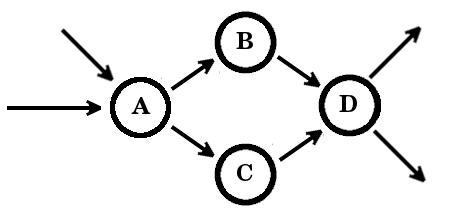}
\caption{\label{last_case} Same as \autoref{sec_case} with four vertices.}
\end{figure}

The last example shown in \autoref{last_case} is a more complex case of branch recombination. Let us assume that no strange particles were produced in the vertices $A$ and $B$ but that strangeness production happened in vertex $C$. $CSR(A)$ and $CSR(B)$ are determined as described in \autoref{ex1} in the case of non-strangeness production. $CSR(C)$ follows the \autoref{importance_calc} because of strangeness production:
\begin{equation}
CSR(C) = \frac{1}{\text{bias ratio} \times CSR(A)}.
\end{equation}

If strangeness is produced in vertex $D$, $CSR(D)$ also follows the \autoref{importance_calc}:
\begin{align}
CSR(D) & = \frac{1}{\text{bias ratio} \times CSR(A) \times CSR(B) \times CSR(C)} \nonumber \\
      & = \frac{1}{CSR(B)}.
\end{align}


\begin{thebibliography}{9}
\addcontentsline{toc}{chapter}{\hspace{5mm}Bibliography}
        
\bibitem{iaea}
		S. Leray et \textit{al.},
		\href{http://www.jkps.or.kr/journal/view.html?uid=12708&vmd=Full}{J. Korean Phys. Soc. 59, 791-796 (2011)} \\
		\href{http://www-nds.iaea.org/spallations}{IAEA Benchmark of Spallation Models}

\bibitem{abla}
		A. Kelić, M. V. Ricciardi, and K.-H. Schmidt,
		\emph{``Proceedings of Joint ICTP-IAEA Advanced Workshop on Model Codes for Spallation Reactions, ICTP Trieste, Italy, 4-8 February 2008''}, edited by D. Filges, S. Leray, Y. Yariv, et \textit{al.}, (IAEA, Vienna, 2008), pp. 181-221.
		
\bibitem{light.ion2}
		D. Mancusi et al.,
		\href{https://doi.org/10.1103/PhysRevC.90.054602}{Phys. Rev. C 90, 054602 (2014)}.
		
\bibitem{removal}
		D. Mancusi et al.,
		\href{https://doi.org/10.1103/PhysRevC.91.034602}{Phys. Rev. C 91, 034602 (2015)}.
		
\bibitem{jose-remove}
		J. L. Rodr\'iguez-S\'anchez et al.,
		\href{https://doi.org/10.1103/PhysRevC.96.054602}{Phys. Rev. C 96, 054602 (2017)}.

\bibitem{Pedoux}
		S. Pedoux and J. Cugnon,
		\href{https://doi.org/10.1016/j.nuclphysa.2011.07.004}{Nucl. Phys. A 866, 16-36 (2011)}.
		
\bibitem{jc}
		J.C. David et al.,
		\href{https://doi.org/10.1140/epjp/i2018-12079-9}{Eur. Phys. J. Plus 133:253 (2018)}.

\bibitem{fritiof}
		V. Uzhinsky
		\href{https://geant4.web.cern.ch/geant4/results/papers/Fritiof-MC2010.pdf}{(SNA + MC2010) Hitotsubashi Memorial Hall, Tokyo, Japan, October 17-21 (2010)}

\bibitem{bibi}
		J. Hirtz, J.C. David, et al.,
		\href{https://doi.org/10.1140/epjp/i2018-12312-7}{Eur. Phys. J. Plus 133:436 (2018)}.
		
\bibitem{incl4.6}
		A. Boudard et al.,
		\href{https://doi.org/10.1103/PhysRevC.87.014606}{Phys. Rev. C 87, 014606 (2013)}.
		
\bibitem{pdg}
		C. Patrignani et al.,
		\href{http://pdg.lbl.gov/}{Chin. Phys. C 40, 100001 (2016)}.

\bibitem{jose}
	    J. L. Rodr\'iguez-S\'anchez, J.-C. David, J. Hirtz, et al.,
	    \href{https://doi.org/10.1103/PhysRevC.98.021602}{Phys. Rev. C 98, 021602(R) (2018)}

\bibitem{pot-meson}
		V. Metag, M. Nanova, and E. Paryev,
		\href{http://www.sciencedirect.com/science/article/pii/S0146641017300704}{Prog. Part. Nucl. Phys. 97, 199-260 (2017)}.

\bibitem{pot-sigma}
		T. Rijken and H.-J. Schulze,
		\href{https://link.springer.com/article/10.1140\%2Fepja\%2Fi2016-16021-6}{Eur. Phys. J. A 52:21 (2016)}.

\bibitem{tsushima}
		K. Tsushima, A. Sibirtsev, A. Thomas, and G. Li,
		\href{http://journals.aps.org/prc/abstract/10.1103/PhysRevC.59.369}{Phys. Rev. C 59, 369 (1999)},
		\href{http://journals.aps.org/prc/abstract/10.1103/PhysRevC.61.029903}{Erratum: Phys. Rev. C 61, 029903 (2000)}.
		
\bibitem{owen}
        A. Owen,
		\href{https://statweb.stanford.edu/~owen/mc/}{Monte Carlo theory, methods and examples (2013)}

\bibitem{kaos}
		W. Scheinast et al.,
		\href{https://journals.aps.org/prl/abstract/10.1103/PhysRevLett.96.072301}{PRL 96, 072301 (2006)}.

\bibitem{bertini}
		D. Wright and M. Kelsey, 
		\href{https://doi.org/10.1016/j.nima.2015.09.058}{NIM A 804, 175-188 (2015)}

\bibitem{itep}
		 A. Akindinov et al.,
		\href{https://doi.org/10.1134/1.1316808}{JETP Letters, Vol. 72, 100 (2000)}.
		
\bibitem{anke}
		M. Büscher, V. Koptev, M. Nekipelov, et al.,
		\href{https://link.springer.com/article/10.1140/epja/i2004-10036-6}{Eur. Phys. J. A 22:301 (2004)}.
		
\bibitem{laqgsm}
		N. Mokhov, K. Gudima, and S. Striganov.,
		\href{https://arxiv.org/abs/1409.1086}{arXiv:1409.1086 [nucl-ex]}.

\bibitem{lbl}
		S. Schnetzer, R. Lombard, et al.,
		\href{https://journals.aps.org/prc/abstract/10.1103/PhysRevC.40.640}{Phys. Rev. C 40, 640 (1990)}.
		\href{https://journals.aps.org/prc/abstract/10.1103/PhysRevC.41.1320}{Erratum: Phys. Rev. C 41, 1320 (1990)}.
		
\bibitem{hades_l}
		HADES Collaboration,
		\href{https://link.springer.com/article/10.1140/epja/i2014-14081-2}{Eur. Phys. J. A 50:81 (2014)}.
		
\bibitem{hades_k}
		HADES Collaboration,
		\href{https://journals.aps.org/prc/abstract/10.1103/PhysRevC.90.054906}{Phys. Rev. C 90, 054906 (2014)}.

\bibitem{UrQMD}
		S. Bass et al.,
		\href{https://arxiv.org/abs/nucl-th/9803035}{arXiv:nucl-th/9803035v2}.

\bibitem{gibuu}
		O. Buss et al.,
		\href{http://dx.doi.org/10.1016/j.physrep.2011.12.001}{Physics Reports 512, 1-124 (2012)}
		
\bibitem{herbach}
	    C.-M. Herbach et al.,
	    \href{https://doi.org/10.1016/j.nima.2006.02.033}{ NIM A 562, 729 (2006)}

\bibitem{fopi}
		M. Benabderrahmane et al.,
		\href{https://journals.aps.org/prl/abstract/10.1103/PhysRevLett.102.182501}{PRL 102, 182501 (2009)}.

\bibitem{qmc}
		K. Tsushima, A. Sibirtsev, and A. Thomas,
		\href{https://journals.aps.org/prc/abstract/10.1103/PhysRevC.62.064904}{Phys. Rev. C 62, 064904 (2000)}.

\bibitem{hsd1}
		W. Cassing et al.,
		\href{https://www.sciencedirect.com/science/article/pii/S0370157398000283?via\%3Dihub}{Phys. Rep. 308, 65 (1999)},
		
\bibitem{hsd2}
		W. Cassing et al.,
		\href{https://www.sciencedirect.com/science/article/pii/S0375947496004617?via\%3Dihub}{Nucl. Phys.A 614, 415 (1997)}

\bibitem{e802}
		E-802 Collaboration,
		\href{https://journals.aps.org/prd/abstract/10.1103/PhysRevD.45.3906}{Phys. Rev. D 45, 3906 (1992)}.

\bibitem{linp}
		V. Koptev et al.,
		\href{http://www.jetp.ac.ru/cgi-bin/e/index/e/67/11/p2177?a=list}{Zh. Eksp. Teor. Fiz. 94,1-14 (1988)}.

\bibitem{geant4}
		Geant4 Collaboration,
		\href{http://geant4-userdoc.web.cern.ch/geant4-userdoc/UsersGuides/PhysicsReferenceManual/fo/PhysicsReferenceManual.pdf}{Geant4 Physics Reference Manual Release 10.5}.
		
\bibitem{HypHI}
        T. Saito et al.,
        \href{https://doi.org/10.1007/978-3-540-76367-3_36}{Proceedings of the IX International Conference on Hypernuclear and Strange Particle Physics. Springer, Berlin, Heidelberg (2007)}

\bibitem{Panda}
        J. Messchendorp,
 	    \href{https://doi.org/10.7566/JPSCP.13.010016}{arXiv:1610.02804 [nucl-ex]}
 	
\bibitem{CBM}
        V. Friese,
        \href{https://doi.org/10.22323/1.047.0056}{Proceedings of Science 47 (2008))}

\end{thebibliography}
\end{document}